\documentclass[12pt,reqno]{amsart}
\usepackage[latin9]{inputenc}
\usepackage{geometry}
\usepackage{float}
\usepackage{textcomp}
\usepackage{mathtools}
\usepackage{amsbsy}
\usepackage{amstext}
\usepackage{amsthm}
\usepackage{amssymb}
\usepackage{mathdots}
\usepackage{graphicx}
\usepackage{esint}
\usepackage[unicode=true,
 bookmarks=true,bookmarksnumbered=true,bookmarksopen=true,bookmarksopenlevel=1,
 breaklinks=false,pdfborder={0 0 1},backref=section,colorlinks=false]
 {hyperref}
\hypersetup{pdftitle={Super-exponential Amplification of Wavepacket Propagation in Traveling Wave Tubes},
 pdfauthor={Kasra Rouhi, Filippo Capolino, Alexander Figotin},
 hypertex,bookmarksnumbered=true}

\makeatletter

\newcommand{\lyxmathsym}[1]{\ifmmode\begingroup\def\b@ld{bold}
  \text{\ifx\math@version\b@ld\bfseries\fi#1}\endgroup\else#1\fi}

\DeclareTextSymbolDefault{\textquotedbl}{T1}
\providecommand{\tabularnewline}{\\}

\numberwithin{equation}{section}
\numberwithin{figure}{section}
\theoremstyle{remark}
\newtheorem*{rem*}{\protect\remarkname}

\usepackage{chngcntr}
\usepackage{amsthm}
\counterwithout{figure}{section}

\makeatother

\providecommand{\remarkname}{Remark}

\begin{document}
\title{Super-exponential Amplification of Wavepacket Propagation in Traveling
Wave Tubes}
\author{Kasra Rouhi}
\author{Filippo Capolino}
\author{Alexander Figotin}
\address{University of California at Irvine, CA 92967.}
\email{afigotin@uci.edu.}
\begin{abstract}
We analyze wavepacket propagation in traveling wave tubes (TWTs) analytically
and numerically. TWT design in essence comprises a pencil-like electron
beam in vacuum interacting with an electromagnetic wave guided by
a slow-wave structure (SWS). In our study, the electron beam is represented
by a one-dimensional electron flow and the SWS is represented by an
equivalent transmission line model. The analytical considerations
are based on the Lagrangian field theory for TWTs. Mathematical analysis
of wavepacket propagation in one-dimensional space is based on the
relevant Euler-Lagrange equations which are second-order differential
equations in both time and space. Wavepacket propagation analysis
is not simple and we develop a numerically efficient algorithm to
perform the analysis efficiently. In particular, when the initial
pulse has a Gaussian shape at the input port, it acquires non-Gaussian
features as it propagates through the TWT. These features include:
(i) super-exponential (faster than exponential) amplification, (ii)
shift of the pulse frequency spectrum toward higher frequencies, and
(iii) change in the shape of the pulse that becomes particularly pronounced
when the pulse frequency band contains a transitional point from stability
to instability.
\end{abstract}

\maketitle

\section{Introduction}

Traveling-wave tubes (TWTs) are widely used in telecommunications,
radar, and high-resolution imaging \cite{shiffler1991high,denisov1998gyrotron,benford2007high,wong2020recent}.
These devices utilize the kinetic energy of a beam of electrons in
vacuum to amplify radio frequency (RF) signals \cite{tsimring2006electron,gilmour2011klystrons}.
J. R. Pierce introduced a simple and effective model that accounts
for (i) transfer of energy from an electron beam to an electromagnetic
wave, and (ii) signal amplification as it propagates through a TWT
\cite{pierce1947theory,pierce1949doulble,pierce1950traveling,pierce1951waves}.
The Pierce model was extensively studied and extended by many authors
\cite{friedman1951amplification,sturrock1958kinematics,kino1960parametric,tamma2014extension}.
Also, in \cite{figotin2013multi,figotin2020analytic}, the analytical
Lagrangian field theory for TWT was constructed to generalize the
original Pierce model. In this model, a system of multi-transmission
line can be coupled to the multi-stream electron beam. Moreover, the
original Pierce theory without considering the space charge effect
\cite{pierce1947theory} can be viewed as a high-frequency approximation
of the Lagrangian field theory \cite[Chapter 29]{figotin2020analytic};
however, when considering the space charge effect as in \cite{pierce1950traveling},
Pierce theory leads to the same results of the Lagrangian field theory,
as demonstrated in the Appendix of \cite{rouhi2021exceptional}.

The subject of pulse amplification is an important topic in numerous
applications such as wideband RF communications, impulse radar, remote
sensing, imaging, spectroscopy, and time domain characterization of
devices. In particular, TWTs have been employed to generate nanosecond
carrier pulses in \cite{klute1951pulse,beck1955microwave,miyauchi1963traveling}
which are useful in radar applications. Furthermore, in \cite{skolnik1990introduction},
the authors show that pulse radars have greater spectral efficiency
than conventional narrow-band radars. In \cite{gritsunov2006propagation},
a technique for the simulation of wideband pulses excited in a regular
dispersive delay line by an electron beam is proposed. Chernin et
al. developed a three-dimensional multi-frequency large signal model
of the beam-electromagnetic wave interaction in a helix TWT, in which
both forward and backward synchronous space harmonics are included
\cite{chernin2001three}. Also, Converse et al. proposed a new one-dimensional
time domain model to be used in the analysis of the pulse response
of the helix TWT \cite{converse2004impulseI,converse2004impulseII}.
Their model incorporates waveguide dispersion and uses a nonlinear
time-domain method to analyze the response of a wideband helix TWT
to an input Gaussian pulse. Setayesh and Abrishamian later developed
a variation of the Converse et al. model \cite{setayesh2017pawaic,setayesh2018new},
extending their pseudo-spectral method to an arbitrary order of accuracy
in both time and space derivatives. Recently, Aliane et al. analyzed
short-pulse amplification in TWTs using the Hamiltonian discrete model
\cite{aliane2021many}. Beyond technological applications, short pulses
are also valuable for probing the TWT dispersion relation and gain.

\begin{figure}
\centering{}\includegraphics[width=1\textwidth]{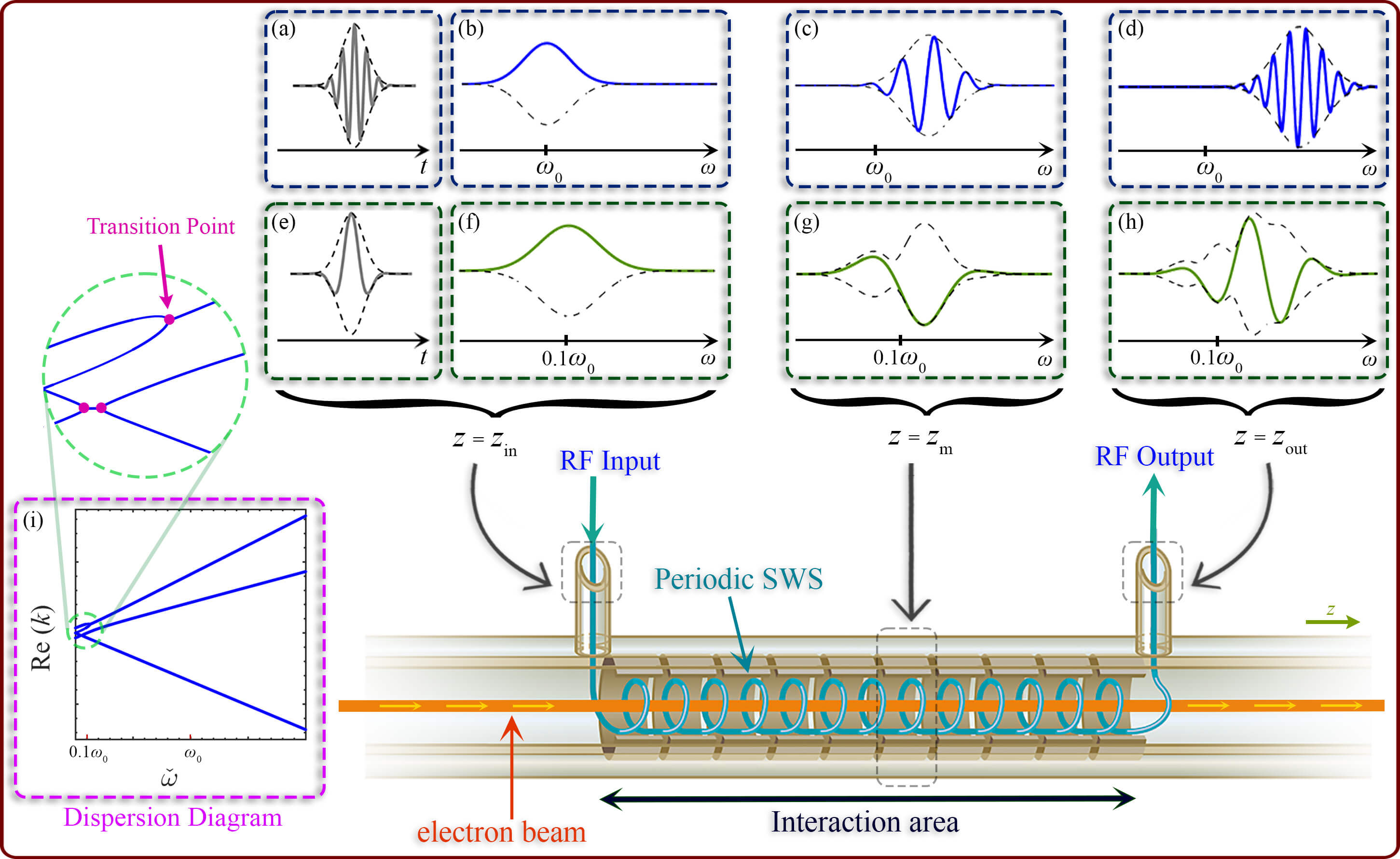}
\caption{The plot illustrates wavepacket propagation in a TWT. The initial
Gaussian pulse enters the RF input port on the left ($z=z_{\mathrm{in}}$),
and the amplified output pulse is extracted from the RF output port
on the right ($z=z_{\mathrm{out}}$). (a)-(d) show wavepacket propagation
when its frequency band is in the constant hot mode complex velocity
region (far from the point of transition between stability and instability).
Instead, (e)-(h) show wavepacket propagation when its frequency band
is close to or covers the transition point. (i) The real part of the
complex-valued wavenumbers of the hot modes which pink dots in the
magnified plot show the transition points. When the operating frequency
is in the region where the complex velocity of the hot modes is more
or less constant (i.e., the dispersionless region), we notice interesting
phenomena during propagation in space such as (i) super-exponential
amplification, (ii) shifts in the center frequency and (iii) an increase
in the number of local peaks in the wavepacket spectrum within the
wavepacket bandwidth. However, if the wavepacket frequency bandwidth
contains a transition point, the wavepacket shape is significantly
altered as it propagates.\label{fig: PulseProp}}
\end{figure}

In this paper, we study pulse propagation and amplification in TWTs
and develop a numerically efficient algorithm to detect nontrivial
features during propagation. In a nutshell, we apply the Lagrangian
field theory for TWTs developed in \cite{figotin2013multi,figotin2020analytic}
to analyze pulse propagation. We conduct our studies initially in
the frequency domain and demonstrate the efficiency of our method
through numerical simulations. According to the results, the propagating
wave is amplified, and the frequency content of the wavepacket is
shifted to higher frequencies when the wavepacket bandwidth is in
the frequency range where the complex-valued wavenumber of the amplifying
hot mode is more or less dispersionless. In addition to that, the
number of local peaks in the wavepacket spectrum increases when the
wavepacket moves away from the input port. An example of these two
phenomena is illustrated in Figures \ref{fig: PulseProp}(b)-(d).
It will be demonstrated that as long as the operation frequency is
in the dispersionless region, the computed results using our developed
``exact method'' and ``approximate method'' show good agreement
(see Section \ref{sec:WavePacketPropagation}). In the approximate
method, we consider constant complex velocity for the \textquotedblleft hot
modes\textquotedblright , which serves as a benchmark (\textquotedblleft hot
modes\textquotedblright{} or ``TWT modes'' are those modes that
account for the interaction of the electromagnetic wave and the charge
waves, and complex velocity refers to the complex-valued analytic
continuation of the phase velocity of the hot mode $u=\omega/k$).
The approximate method fails to yield reliable predictions when the
wavepacket frequency spectrum composition includes transition points.
This occurs due to an abrupt change in the complex velocity of the
hot modes at transition points and the wavepacket shape and frequency
content are altered during propagation as shown in Figures \ref{fig: PulseProp}(f)-(h).
Thus, when the complex velocity of the hot mode varies significantly
with frequency, like when the pulse bandwidth contains a transition
point, the exact method we propose is the only alternative. Our developed
method and its numerical implementation offer valuable insights into
pulse propagation and amplification in TWTs, enabling the design of
devices that operate effectively with pulsed RF input signals.

This article is organized as follows. Section \ref{sec:Significant Results}
concisely highlights the key results and nontrivial phenomena during
wavepacket propagation in TWTs. In Section \ref{sec:Lagrangian-Basics},
we briefly review the Lagrangian field theory employed in the analysis
of TWTs. Section \ref{sec:WavePacketPropagation} discusses the analytical
approach and associated approximations for modeling wavepacket propagation
in TWTs. In Section \ref{sec:Results}, we show the effectiveness
of the developed analytical method and its approximation using several
examples, including one based on a realistic TWT design. Finally,
we conclude the paper in Section \ref{sec:Conclusions}.

\section{Statement of The Main Results\label{sec:Significant Results}}

We use an analytical model to investigate key features of wavepacket
propagation in TWTs. We find that if the pulse frequency composition
involves exponentially growing modes, the pulse shape and frequency
content change significantly as it propagates through the TWT. Here,
we briefly highlight our main results and achievements, leaving explanations,
technical details, and numerical examples for the following sections.

Our main statements regarding pulse propagation in a typical TWT can
be summarized as follows:
\begin{enumerate}
\item Detectable super-exponential amplification (see Remark \ref{Super-exponential})
of the wavepacket as it propagates through the TWT (see Figures \ref{fig: PulseProp}(b)-(d));
\item Noticeable shift of the pulse frequency band toward higher frequencies
(see Figures \ref{fig: PulseProp}(c) and (d));
\item Noticeable increase in the number of local peaks in the wavepacket
spectrum in the pulse bandwidth as it propagates through the TWT (see
Figures \ref{fig: PulseProp}(c) and (d));
\item Significant distortion in the pulse shape occurs when the pulse frequency
bandwidth contains the frequency of a transition point separating
stability from instability (see Figures \ref{fig: PulseProp}(f)-(h)).
\end{enumerate}
The above-mentioned results are illustrated in Figure \ref{fig: PulseProp}.
In the first row, the center frequency of the input pulse is located
far away from the transition frequency of the TWT dispersion diagram
(see Figures \ref{fig: PulseProp}(a) and (b)). The hot mode phase
velocity is constant in this region, and the propagated wavepacket
can also be calculated via the approximate method presented in \cite[Chapter 16]{figotin2020analytic}.
In this case, the wavepacket center frequency is shifted as the pulse
propagates through the TWT (see Figures \ref{fig: PulseProp}(c) and
(d)). In addition, the number of local peaks in the wavepacket spectrum
in the pulse bandwidth is gradually increased while the wavepacket
travels toward the output port. Also, the wavepacket amplitude is
amplified rapidly, representing the most significant feature of TWTs
in the amplification regime. In the second row of Figure \ref{fig: PulseProp},
the Gaussian pulse with the low center frequency enters the input
port of the TWT (see Figures \ref{fig: PulseProp}(e) and (f)). The
center frequency of the input pulse is located near the transition
point, i.e., the bifurcation point in the dispersion diagram. The
complex-valued analytic continuation of the hot mode phase velocity
near transition points is strongly dependent on frequency. So, every
spectral component of the pulse is amplified with a different amplitude
and phase, leading to distortion in the propagated wavepacket. As
a result, the extracted wavepacket shapes at the middle of TWT and
at the output port do not resemble the original Gaussian pulse shape
(see Figures \ref{fig: PulseProp}(g) and (h)). Consequently, when
the wavepacket frequency bandwidth contains the transition point frequency,
we observe significant alterations in the wavepacket envelope. Detailed
analysis of these nontrivial effects and phenomena requires more accurate
investigation, which is available in Section \ref{sec:WavePacketPropagation}
with different examples in Section \ref{sec:Results}.

\section{Review of The Lagrangian Field Theory For TWT\label{sec:Lagrangian-Basics}}

We briefly review the analytical model used in this paper for studying
TWTs. As a pioneer in TWT modeling, Pierce developed a mathematical
model \cite{pierce1950traveling}. This model can be considered the
simplest one that accounts for electromagnetic wave amplification
in the TWT \cite{tsimring2006electron,gilmour2011klystrons}. The
Pierce model, also known as the 4-wave theory of a TWT, is a one-dimensional
linear theory in which the slow wave structure (SWS) is represented
by a transmission line assumed to be homogeneous \cite{pierce1947theory,pierce1950traveling}.
Also, the 3-wave small-signal theory, which laid the foundation for
TWT design, can be viewed as an approximation of the 4-wave theory
\cite{pierce1950traveling}.

In this paper, we use an analytical model based on the Lagrangian
field framework developed in \cite{figotin2013multi,figotin2020analytic}.
The Lagrangian field theory also allows for modeling more complex
SWSs than Pierce's simple one by involving more than one guided electromagnetic
wave and the multi-stream beam \cite{figotin2013multi,figotin2020analytic}.
As shown in \cite{rouhi2021exceptional}, the equivalent transmission
line equations that one would get directly from generalizing the Pierce
model agree with the analytical Lagrangian field theory.

\subsection{Fundamental equations\label{subsec:Fundamental-equations}}

\begin{figure}
\centering{}\includegraphics[width=0.45\textwidth]{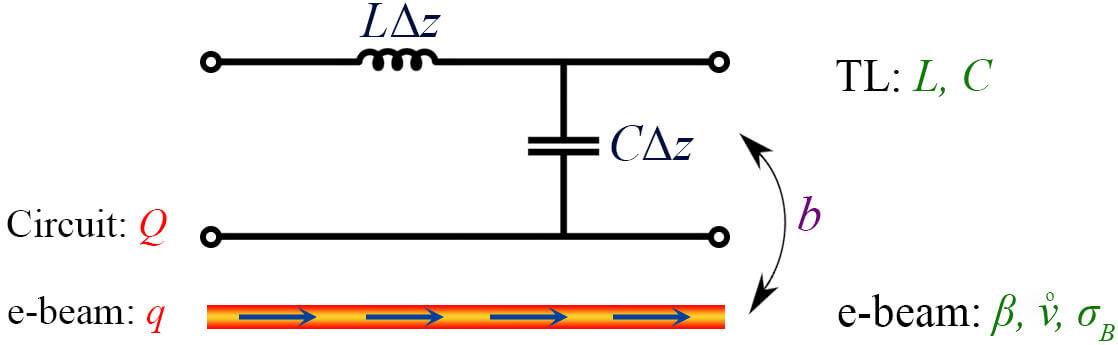}
\caption{Equivalent transmission line representation of the SWS where the equivalent
transmission line is represented conceptually as a circuit, with $C$
and $L$ being the per-unit-length capacitance and inductance, respectively,
and \textit{$b$} represents the coupling strength with the charge
waves.\label{fig: Model}}
\end{figure}

We assume the electron stream to be confined by an external static
magnetic field to an infinitely long cylinder along the $z$ direction
\cite{nishihara1970measurement}. The area of the cross-section $\sigma_{\mathrm{B}}$
of the electron stream is supposed to be small enough to ignore transverse
variations in relevant physical quantities. The plasma frequency of
the corresponding electron stream is given by \cite{bohm1949theory,lampert1956plasma}

\begin{equation}
\omega_{\mathrm{p}}^{2}=\frac{4\pi\mathring{n}\mathrm{e}^{2}}{m},
\end{equation}
where $\mathring{n}$ denotes electron density, $-\mathrm{e}$ is
the electron charge, and $m$ is the electron rest mass. The electron
stream steady velocity is denoted by $\mathring{v}$, and another
key parameter related to the electron stream is the stream intensity,
which is defined as

\begin{equation}
\beta=\frac{\sigma_{\mathrm{B}}}{4\pi}R_{\mathrm{sc}}^{2}\omega_{\mathrm{p}}^{2},
\end{equation}
where $R_{\mathrm{sc}}$ is the so-called plasma frequency reduction
factor that accounts phenomenologically for the finite dimensions
of the electron stream cylinder and geometric features of the SWS
\cite{branch1955plasma}. The physical dimension of the electron stream
intensity is the square of velocity. A larger stream intensity value
is associated with a larger value of electron density or a larger
electron stream cross-sectional area. The electron stream steady velocity
$\mathring{v}$, and electron stream intensity $\beta$, play a significant
role in defining TWT properties, so we combine them into a set of
electron stream significant parameters as in \cite{figotin2020analytic}.
We use the stream charge to build up the electron stream interaction
with the electromagnetic wave, defined as the time integral of the
corresponding electron stream current $i\left(t,z\right)$, as $q\left(t,z\right)=\intop_{t_{0}}^{t}i\left(t',z\right)\:\mathrm{d}t'$.
The variable $q\left(t,z\right)$ represents the amount of charge
that has traversed the electron stream cross-section, at point $z$,
from the initial time of $t_{0}$ to time $t$. Electromagnetic propagation
in the SWS is modeled as a single equivalent transmission line, which
describes the electromagnetic modal propagation in the SWS, based
on the equivalent transmission line model shown in Figure \ref{fig: Model}.
Here, the distributed per-unit-length series inductance $L$ and shunt
capacitance $C$ are used to describe the electromagnetic properties
of the equivalent transmission line \cite{brillouin1949traveling,haus1955noise,milton2006electromagnetic,paul2007analysis}.
In the well-known definition of equivalent transmission line characteristics,
we use voltage $V(t,z)$ and current $I(t,z)$, alongside the charge
definition $Q\left(t,z\right)=\intop_{t_{0}}^{t}I\left(t',z\right)\:\mathrm{d}t'$.
This value is a primary parameter for describing the equivalent transmission
line, which indicates the amount of charge that has crossed the cross-section
of the equivalent transmission line from the initial time of $t_{0}$
to time $t$. Note that $q\left(t,z\right)$ and $Q\left(t,z\right)$
are purely real in the Lagrangian equations. Also, cutoff conditions
could be modeled by resonant series and shunt reactive elements in
the transmission line equivalent circuit model \cite{kino1962circuit,singh1983equivalent,miano2001transmission}.
However, for simplicity, in this paper we assume we work not close
to a waveguide cutoff frequency and hence we ignore this feature.
Accordingly, using the transmission line formalism, the phase velocity
of the electromagnetic wave is calculated as $w=1/\sqrt{LC}$. The
coupling strength between the electron stream and the electromagnetic
guided wave in the SWS is represented by the parameter $b$ (denoted
by $a$ in \cite{tamma2014extension,rouhi2021exceptional,rouhi2024parametric}).
The term $b$ describes how the electron stream couples to the electromagnetic
wave where the representation of the coupling between an electron
stream and electromagnetic wave guided by SWS goes back to Ramo \cite{ramo1939currents}.
The value of $b=0$ indicates that the electron stream is not coupled
to the guided electromagnetic wave.

Then, following \cite{figotin2013multi,figotin2020analytic}, we introduce
the TWT system Lagrangian $\mathcal{L}_{\mathrm{TB}}$ as

\begin{equation}
\mathcal{L}_{\mathrm{TB}}=\mathcal{L}_{\mathrm{Tb}}+\mathcal{L}_{\mathrm{B}},
\end{equation}
where the Lagrangian components $\mathcal{L}_{\mathrm{Tb}}$ and $\mathcal{L}_{\mathrm{B}}$
are associated with the electromagnetic wave and the electron stream
respectively and are defined as follows \cite{figotin2013multi,figotin2020analytic}

\begin{equation}
\mathcal{L}_{\mathrm{Tb}}=\frac{L}{2}\left(\partial_{t}Q\right)^{2}-\frac{1}{2C}\left(\partial_{z}Q+b\partial_{z}q\right)^{2},
\end{equation}

\begin{equation}
\mathcal{L}_{\mathrm{B}}=\frac{1}{2\beta}\left(\partial_{t}q+\mathring{v}\partial_{z}q\right)^{2}-\frac{2\pi}{\sigma_{\mathrm{B}}}q^{2}.
\end{equation}
The symbols $\partial_{t}$ and $\partial_{z}$ represent the partial
derivative with respect to time $t$ and space $z$, respectively.
Also, the space-charge debunching effects are considered by the term
$2\pi q^{2}/\sigma_{\mathrm{B}}$ in the above equation. For the TWT
system, we find the following Euler-Lagrange equations associated
with the Lagrangian \cite{figotin2013multi,figotin2020analytic}

\begin{equation}
L\partial_{t}^{2}Q-\partial_{z}\left[C^{-1}\left(\partial_{z}Q+b\partial_{z}q\right)\right]=0,\label{eq: Euler-Lagrange1}
\end{equation}

\begin{equation}
\frac{1}{\beta}\left(\partial_{t}+\mathring{v}\partial_{z}\right)^{2}q+\frac{4\pi}{\sigma_{\mathrm{B}}}q-b\partial_{z}\left[C^{-1}\left(\partial_{z}Q+b\partial_{z}q\right)\right]=0.\label{eq: Euler-Lagrange2}
\end{equation}

In the following, we look at the spectrum content of $Q$ and $q$,
and we use the standard complex-signal notation by assuming that all
the quantities are proportional to $\exp\left(\mathrm{i}\omega t\right)$;
therefore; the time derivative leads to multiplying by $\mathrm{i}\omega$
for every spectrum component. We introduce the TWT principal parameter
$\gamma$, which is assumed to be frequency-independent

\begin{equation}
\gamma=\frac{b^{2}}{C}\beta.\label{eq: Gamma}
\end{equation}
According to this definition, we have a larger value for the TWT principal
parameter $\gamma$ when a higher coupling strength coefficient $b$
or a stream intensity $\beta$ (i.e., larger electron density) are
considered. In addition, if we select the smaller shunt capacitance
per-unit-length $C$, we obtain a larger value of the TWT principal
parameter $\gamma$. To obtain additional physical insights, we use
the convenient dimensionless version of the TWT principal parameter
$\check{\gamma}$, defined as

\begin{equation}
\check{\gamma}=\frac{\gamma}{\mathring{v}^{2}}.\label{eq: Gamma_Norm}
\end{equation}

Then, we define the system's state vector as $\boldsymbol{\Psi}\left(z\right)\equiv\left[Q,\partial_{z}Q,q,\partial_{z}q\right]^{\mathrm{T}}$,
where $\mathrm{T}$ indicates transpose operator. The state vector
consists of a combination of charge in the electron stream, charge
in the equivalent transmission line and their first-order spatial
derivatives. Finally, the system evolution along the $z$ direction
is described by the differential equation,

\begin{equation}
\partial_{z}\boldsymbol{\Psi}\left(z\right)=-\mathrm{i}\mathbf{\underline{M}}\boldsymbol{\Psi}\left(z\right),\label{eq: Differential Equation}
\end{equation}
where $\mathbf{\underline{M}}$ is a $4\times4$ TWT matrix and is
given by
\begin{flushright}
\begin{equation}
\mathbf{\underline{M}}=\left[\begin{array}{rrrr}
0 & \mathrm{i} & 0 & 0\\
\mathrm{i}\left(\check{\gamma}-1\right)\frac{\omega^{2}}{w^{2}} & 0 & \mathrm{i}\frac{b}{\mathring{v}^{2}}\left(R_{\mathrm{sc}}^{2}\omega_{\mathrm{p}}^{2}-\omega^{2}\right) & -\frac{2\omega}{\mathring{v}}b\\
0 & 0 & 0 & \mathrm{i}\\
-\mathrm{i}\check{\gamma}\frac{1}{b}\frac{\omega^{2}}{w^{2}} & 0 & -\mathrm{i}\frac{1}{\mathring{v}^{2}}\left(R_{\mathrm{sc}}^{2}\omega_{\mathrm{p}}^{2}-\omega^{2}\right) & \frac{2\omega}{\mathring{v}}
\end{array}\right].\label{eq: TWT Matrix}
\end{equation}
\par\end{flushright}

Note that here, we assume the equivalent transmission line to be uniform,
i.e., $z$ invariant; therefore, $\mathbf{\underline{M}}$ is invariant
in the $z$ direction, but a further generalization to treat nonuniform
SWSs is also possible.

\subsection{Modal dispersion relation}

We assume a state vector $z$-dependence of the form $\boldsymbol{\Psi}\left(z\right)\propto\exp\left(\mathrm{i}\omega t-\mathrm{i}kz\right)$,
where $k$ is the complex-valued wavenumber of a hot mode that accounts
for the interaction of the electromagnetic wave in the SWS with the
beam's charge waves. Then, the four hot eigenmodes are obtained by
solving the eigenvalue problem $k\boldsymbol{\Psi}\left(z\right)=\mathbf{\underline{M}}\boldsymbol{\Psi}\left(z\right)$.
The modal dispersion relation or characteristic equation is given
by

\begin{equation}
D\left(\omega,k\right)=\det\left(\mathbf{\underline{M}}-k\mathbf{\underline{I}}\right)=0.
\end{equation}
After some mathematical calculation and by using the system matrix
in Equation (\ref{eq: TWT Matrix}), the dispersion relation is expressed
by

\begin{equation}
D\left(\omega,k\right)\equiv k^{4}-2k^{3}\frac{\omega}{\mathring{v}}+k^{2}\left(\frac{\omega^{2}}{\mathring{v}^{2}}-\frac{\omega^{2}}{w^{2}}+\frac{\check{\gamma}\omega^{2}}{w^{2}}-\frac{R_{\mathrm{sc}}^{2}\omega_{\mathrm{p}}^{2}}{\mathring{v}^{2}}\right)+2k\frac{\omega^{3}}{w^{2}\mathring{v}}+\omega^{2}\frac{R_{\mathrm{sc}}^{2}\omega_{\mathrm{p}}^{2}-\omega^{2}}{w^{2}\mathring{v}^{2}}=0.\label{eq: Dispersion}
\end{equation}

The solution of the dispersion relation leads to four complex modal
wavenumbers that describe the four hot modes in the electromagnetic
and electron stream interactive system. Since all coefficients in
the characteristic equation are purely real, modal wavenumber solutions
are either purely real or complex conjugates. If we simplify the dispersion
relation by using the complex-valued analytic continuation of the
phase velocity of the hot mode $u=\omega/k$ (in the following simply
referred to as complex velocity), we obtain a simplified form of the
TWT characteristic equation \cite{figotin2020analytic}

\begin{equation}
\frac{\gamma}{w^{2}-u^{2}}+\frac{\left(u-\mathring{v}\right)^{2}}{u^{2}}=\frac{1}{\check{\omega}^{2}},\qquad\check{\omega}=\frac{\omega}{R_{\mathrm{\mathrm{sc}}}\omega_{\mathrm{p}}},\label{eq: DispersionShort}
\end{equation}
where $\check{\omega}$ is the normalized (dimensionless) frequency
which is real-valued in our analysis. We have concisely written the
dispersion equation as a function of the three main physical parameters:
(i) the electron stream steady velocity $\mathring{v}$; (ii) the
cold electromagnetic wave modal phase velocity $w$; and (iii) the
TWT principal parameter $\gamma$. A convenient dimensionless form
of the same dispersion relation is expressed by

\begin{equation}
\frac{\check{\gamma}}{\check{w}^{2}-\check{u}^{2}}+\frac{\left(\check{u}-1\right)^{2}}{\check{u}^{2}}=\frac{1}{\check{\omega}^{2}},\qquad\check{u}=\frac{u}{\mathring{v}},\qquad\check{w}=\frac{w}{\mathring{v}},\label{eq: DispersionShortDimensionless}
\end{equation}
where $\check{u}$ and $\check{w}$ are dimensionless parameters.
The Euler-Lagrange relations in Equations (\ref{eq: Euler-Lagrange1})
and (\ref{eq: Euler-Lagrange2}) and the system of first-order differential
equation in Equation (\ref{eq: Differential Equation}) are written
in the centimeter-gram-second system (CGS-Gaussian system). However,
the dispersion relation is written using dimensionless parameters
and therefore SI units could also be used. The table for CGS to SI
transformation is provided in Appendix \ref{sec:AppendixCGSSI}. The
formalism translation between the Lagrangian model parameters used
in this framework and the parameters used in the Pierce model is listed
in Appendix \ref{sec:AppendixPierce}, and more details are in \cite{rouhi2021exceptional}.

\section{Wavepacket Propagation in Traveling-Wave Tube\label{sec:WavePacketPropagation}}

An analytical approach to studying wavepacket propagation in TWTs
is presented in this section. We focus on Gaussian pulse propagation
in TWTs but the analysis can be extended to other forms of wavepackets.
We use analytical formulas based on the Lagrangian field theory explained
in Section \ref{sec:Lagrangian-Basics} to describe wavepacket propagation.
Then, we calculate the approximate propagated wavepacket under specific
assumptions. The results are compared to those using the transfer
matrix approach that takes into account all the four hot modes in
the TWT.

\subsection{Wavepacket propagation and amplification under the assumption that
$u$ is constant (approximate method)\label{subsec:Approximate-method}}

Any wavepacket propagating in a dispersive medium can be represented
as a linear composition of its spectral constituents of different
frequencies. Ideally, a wavepacket propagates through a medium without
distortions if the relevant modal dispersion relation is linear, i.e.,
the complex velocity of each hot mode $u=\omega/k$ is constant and
does not depend on frequency. In this subsection, we consider only
the growing hot mode that amplifies the wavepacket and neglect the
frequency variation of its phase velocity in the frequency domain
of interest. We assume that for this single dispersionless growing
hot mode $u\left(\omega\right)=\overline{u}$, where $\overline{u}$
is constant with imaginary part $\Im\left\{ \overline{u}\right\} <0$
that leads to amplification (see Figure \ref{fig: ExampleDispersion}(e)
and (f)). The dispersionless assumption for this mode is valid only
when the operating frequency is far enough from a transition point
in the wavenumber dispersion diagram (see wavenumber dispersion diagram
in Figures \ref{fig: PulseProp}(i) and \ref{fig: ExampleDispersion}(e)).

A Gaussian wavepacket that represents the forward electromagnetic
wave in the TWT is described by the complex charge representation,

\begin{equation}
Q\left(t,z\right)=\frac{A}{2\pi}\exp\left(\mathrm{i}\omega_{0}\left(t-\frac{z}{\overline{u}}\right)-\frac{1}{\tau_{\mathrm{w}}^{2}}\left(t-\frac{z}{\overline{u}}\right)^{2}\right),\label{eq: Gaussian Pulse}
\end{equation}
where $A$ is the pulse amplitude (in unit of charge), $\omega_{0}$
indicates the pulse center angular frequency, $\tau_{\mathrm{w}}$
show the pulse time constant. The wavepacket frequency information
is obtained by applying the Fourier transform as defined in \cite[Chapter 4]{oppenheim1996signals},
leading to

\begin{equation}
\widetilde{Q}\left(\omega,z\right)=\frac{A\tau_{\mathrm{w}}}{2\sqrt{\pi}}\exp\left(-\frac{\tau_{\mathrm{w}}^{2}\left(\omega-\omega_{0}\right)^{2}}{4}-\mathrm{i}\omega\frac{z}{\overline{u}}\right).
\end{equation}
The complex velocity of the growing hot mode is calculated in terms
of the power gain factor $\alpha_{0}<0$ that is defined as \cite[Chapter 6.1]{figotin2020analytic}

\begin{equation}
\alpha_{0}=-\frac{\Im\left\{ k\right\} }{\Re\left\{ k\right\} }=\frac{\Im\left\{ \overline{u}\right\} }{\Re\left\{ \overline{u}\right\} }.\label{eq: Alpha0}
\end{equation}
In the above equation, $\Im\left\{ \overline{u}\right\} $ shows the
imaginary part and $\Re\left\{ \overline{u}\right\} $ shows the real
part of the growing hot mode complex velocity. Therefore, we conveniently
rewrite the single dispersionless growing hot mode complex velocity
as \cite[Chapter 16.2]{figotin2020analytic}

\begin{equation}
\overline{u}=\Re\left\{ \overline{u}\right\} \left(1+\mathrm{i}\alpha_{0}\right).
\end{equation}
It is convenient to define another parameter, related to wave energy
velocity \cite[Chapter 16.2]{figotin2020analytic},

\begin{equation}
u_{\mathrm{en}}=\Re\left\{ \overline{u}\right\} \left(1+\alpha_{0}^{2}\right),\label{eq: u_en}
\end{equation}
which is called the pseudo-real part of $\overline{u}$, defined as
$1/u_{\mathrm{en}}$= $\Re\left\{ 1/\overline{u}\right\} $. Notice
that the wave energy velocity $u_{\mathrm{en}}$ is larger than the
real part of the hot mode complex velocity $\Re\left\{ \overline{u}\right\} $
when $\Im\left\{ \overline{u}\right\} \neq0$. The wave energy velocity
$u_{\text{en }}$ can be significantly larger than $\Re\{\overline{u}\}$
if $\Im\{\overline{u}\}$, and consequently the power gain factor
$\alpha_{0}$, is sufficiently large \cite[Chapter 16.2]{figotin2020analytic}.
Then, the exponent of $\widetilde{Q}\left(\omega,z\right)$ is given
by

\begin{equation}
\ln\left(\widetilde{Q}\left(\omega,z\right)\right)=\ln\left(\frac{A\tau_{\mathrm{w}}}{2\sqrt{\pi}}\right)-\frac{\tau_{\mathrm{w}}^{2}\left(\omega-\omega_{0}\right)^{2}}{4}-\mathrm{i}\omega\frac{z}{\overline{u}}.
\end{equation}
We obtain the following expressions for the real and imaginary parts
\cite[Chapter 16.2]{figotin2020analytic},

\begin{equation}
\Re\left\{ \ln\left(\widetilde{Q}\left(\omega,z\right)\right)\right\} =\ln\left(\frac{A\tau_{\mathrm{w}}}{2\sqrt{\pi}}\right)-\frac{\tau_{\mathrm{w}}^{2}\left(\omega-\omega_{0}\right)^{2}}{4}-\frac{\omega\alpha_{0}}{u_{\mathrm{en}}}z,
\end{equation}

\begin{equation}
\Im\left\{ \ln\left(\widetilde{Q}\left(\omega,z\right)\right)\right\} =-\frac{\omega}{u_{\mathrm{en}}}z.
\end{equation}
Next, the real part is rewritten as

\begin{equation}
\Re\left\{ \ln\left(\widetilde{Q}\left(\omega,z\right)\right)\right\} =\ln\left(\frac{A\tau_{\mathrm{w}}}{2\sqrt{\pi}}\right)-\frac{\tau_{\mathrm{w}}^{2}\left(\omega-\omega_{\mathrm{w}}\left(z\right)\right)^{2}}{4}+\frac{\tau_{\mathrm{w}}^{2}\left(\omega_{\mathrm{w}}^{2}\left(z\right)-\omega_{0}^{2}\right)}{4}\label{eq: RealPart}
\end{equation}
\[
=\ln\left(\frac{A\tau_{\mathrm{w}}}{2\sqrt{\pi}}\right)-\frac{\tau_{\mathrm{w}}^{2}\left(\omega-\omega_{\mathrm{w}}\left(z\right)\right)^{2}}{4}+\frac{\tau_{\mathrm{w}}^{2}\omega_{0}^{2}\delta_{\mathrm{w}}\left(z\right)\left(2+\delta_{\mathrm{w}}\left(z\right)\right)}{4},
\]
where

\begin{equation}
\omega_{\mathrm{w}}\left(z\right)=\omega_{0}\left(1+\delta_{\mathrm{w}}\left(z\right)\right),\label{eq: Omega_w}
\end{equation}

\begin{equation}
\delta_{\mathrm{w}}\left(z\right)=\frac{\omega_{\mathrm{w}}\left(z\right)-\omega_{0}}{\omega_{0}}=-\frac{2\alpha_{0}}{u_{\mathrm{en}}\omega_{0}\tau_{\mathrm{w}}^{2}}z.\label{eq: delta_w}
\end{equation}
These two equations define $\omega_{\mathrm{w}}\left(z\right)$ as
the wavepacket shifted angular frequency, and $\delta_{\mathrm{w}}\left(z\right)$
as the wavepacket relative frequency shift. These new quantities demonstrate
that the complex velocity of the hot mode $\overline{u}$ causes the
shift of the wavepacket center angular frequency from $\omega_{0}$
to $\omega_{\mathrm{w}}\left(z\right)$ and the angular frequency
shift is calculated by $\omega_{0}\delta_{\mathrm{w}}\left(z\right)$.
Note that the frequency shift is directly associated to the presence
of the imaginary part of $\overline{u}$. As we observe in Equations
(\ref{eq: Omega_w}) and (\ref{eq: delta_w}), $\omega_{\mathrm{w}}\left(z\right)$
and $\delta_{\mathrm{w}}\left(z\right)$ are position-dependent. Equation
(\ref{eq: delta_w}) implies that the relative frequency deviation
from the center frequency increases linearly when the exponentially
growing wavepacket travels inside the TWT (increasing $z$). The field
amplification exponent or amplification factor appears in $\Re\left\{ \ln\left(\widetilde{Q}\left(\omega,z\right)\right)\right\} $,
and it accounts for the wavepacket amplification \cite[Chapter 16.2]{figotin2020analytic}.
Therefore, the signal amplification is defined as

\begin{equation}
S_{\mathrm{amp}}\left(z\right)=\ln\left(\frac{\widetilde{Q}\left(\omega_{\mathrm{w}}\left(z\right),z\right)}{\widetilde{Q}\left(\omega_{0},0\right)}\right)=\ln\left(\frac{\widetilde{Q}\left(\omega_{\mathrm{w}}\left(z\right),z\right)}{A\tau_{\mathrm{w}}/\left(2\sqrt{\pi}\right)}\right).
\end{equation}
and, according to Equation (\ref{eq: RealPart}), it is equal to

\begin{equation}
S_{\mathrm{amp}}\left(z\right)=\frac{\tau_{\mathrm{w}}^{2}\left(\omega_{\mathrm{w}}^{2}\left(z\right)-\omega_{0}^{2}\right)}{4}=\frac{\tau_{\mathrm{w}}^{2}\omega_{0}^{2}\delta_{\mathrm{w}}\left(z\right)\left(2+\delta_{\mathrm{w}}\left(z\right)\right)}{4},\label{eq: S_amp}
\end{equation}
This parameter shows the ratio between the maximum value of the amplified
wavepacket at the shifted center frequency and the maximum value of
the input Gaussian pulse by assuming the excitation of single dispersionless
growing hot mode. It is clear that the maximum value of the propagated
wavepacket occurs when the second term in Equation (\ref{eq: RealPart})
vanishes since it has a negative value. Thus, the maximum value is
obtained at $\omega=\omega_{\mathrm{w}}\left(z\right)$, which is
the center frequency of the traveled wavepacket. According to Equation
(\ref{eq: S_amp}), an amplification factor is position-dependent
because it is related to the relative frequency shift along $z$.
As already mentioned, in this subsection, we used the charge $\widetilde{Q}\left(\omega,z\right)$
to describe the propagated wavepacket; however, this method can be
applied to other quantities of the wavepacket such as current, electric
fields, etc.
\begin{rem*}[\label{Super-exponential}Super-exponential amplification]
 Generally speaking, super-exponential means more than exponential,
so a function is super-exponential if it grows faster than any exponential
function. According to Equation (\ref{eq: S_amp}), the amplification
factor can be written as

\begin{equation}
S_{\mathrm{amp}}\left(z\right)=C_{1}\delta_{\mathrm{w}}\left(z\right)+C_{2}\delta_{\mathrm{w}}^{2}\left(z\right),
\end{equation}
where $C_{1}$ and $C_{2}$ are constant, system dependent, coefficients.
On the other hand, the wavepacket shifted relative frequency is a
linear function of $z$, i.e., $\delta_{\mathrm{w}}\left(z\right)\propto z$
(see Equation (\ref{eq: delta_w})). As a result, the amplification
occurs super-exponentially or faster than exponential by propagating
through the TWT (increasing $z$) due to the presence of a quadratic
term at the exponent, which results in super-exponential growth.
\end{rem*}

\subsection{Gaussian wavepacket as an excitation pulse (Exact method)\label{subsec:Exact-Method}}

In the frequency domain, the initial boundary condition state vector
at $z=0^{+}$ is

\begin{equation}
\boldsymbol{\Psi}\left(z_{0}\right)=\left[\begin{array}{r}
\left.\widetilde{Q}\left(\omega,z\right)\right|_{z=0^{+}}\\
\left.\partial\widetilde{Q}\left(\omega,z\right)/\partial z\right|_{z=0^{+}}\\
0\\
0
\end{array}\right].
\end{equation}
If we assume the field is excited by a Gaussian pulse at the beginning
of the TWT, the field initial conditions are

\begin{equation}
\left\{ \begin{array}{l}
\left.\widetilde{Q}\left(\omega,z\right)\right|_{z=0^{+}}=\frac{A\tau_{\mathrm{w}}}{2\sqrt{\pi}}\exp\left(-\frac{\tau_{\mathrm{w}}^{2}\left(\omega-\omega_{0}\right)^{2}}{4}\right)\\
\left.\partial\widetilde{Q}\left(\omega,z\right)/\partial z\right|_{z=0^{+}}=-\frac{\mathrm{i}\omega}{\overline{u}}\left.\widetilde{Q}\left(\omega,z\right)\right|_{z=0^{+}}
\end{array}\right.\label{eq: Initial Condition Frequency}
\end{equation}
In the second of the latter equations, we have assumed that the initial
velocity is the one of the hot mode $\overline{u}$. In the frequency
domain, the propagated wavepacket is represented by the state vector
at any arbitrary coordinate $z=z_{1}$ via

\begin{equation}
\boldsymbol{\Psi}\left(z_{1}\right)=\mathbf{\underline{T}}\left(z_{1},z_{0}\right)\boldsymbol{\Psi}\left(z_{0}\right),
\end{equation}
where $\mathbf{\underline{T}}\left(z_{1},z_{0}\right)=\exp\left(-\mathrm{i}\mathbf{\underline{M}}\left(z_{1}-z_{0}\right)\right)$
is the TWT transfer matrix, which transfers the state vector $\boldsymbol{\Psi}\left(z\right)$
between the two points of $z_{0}$ and $z_{1}$ in the TWT. It is
assumed in this calculation that the TWT is fully matched in the output
and there is no reflection from that port.

\section{Results and Discussion\label{sec:Results}}

According to the discussion in Section \ref{sec:Lagrangian-Basics},
we need two sets of parameters to analyze wavepacket propagation.
Firstly, we need the fundamental TWT parameters, and secondly, we
should express the initial pulse at an arbitrary position in the TWT,
i.e., at the RF input port. In Subsections \ref{exa:Example1}-\ref{exa:Example3},
we provide examples that are designed in a special way to highlight
some nontrivial features of pulse propagation in TWTs. Then, in Subsection
\ref{exa:ExampleHelix}, we use some realistic parameters for a design
in the microwave regime.

\begin{figure}
\centering{}\includegraphics[width=1\textwidth]{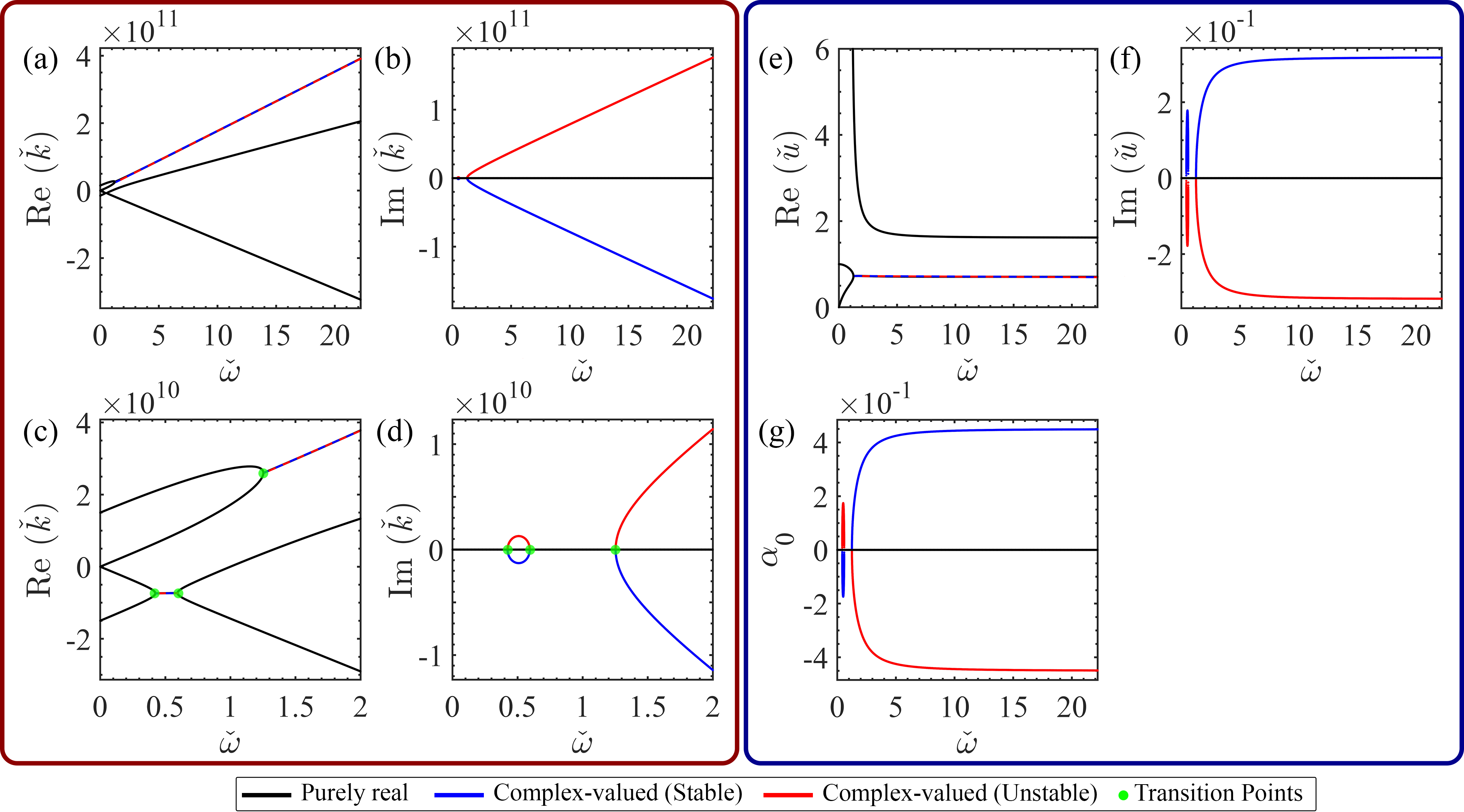}
\caption{Normalized wavenumber $k$ of the hot modes as a function of normalized
angular frequency for the data in Examples (1)-(3) that are provided
in Subsections \ref{exa:Example1}-\ref{exa:Example3}: (a) and (c),
real part; (b) and (d), imaginary part. The bottom plots (c) and (d)
are magnified fragments of the dispersion relation in (a) and (b)
showing more details around the transition points. The green dots
show the transition points separating stability from instability.
(e) and (f) show the real and imaginary parts of the complex velocity
of the hot modes $u=\omega/k$ which are approximately frequency-independent
at the high frequency. (g) Power gain factor $\alpha_{0}$ which is
negative for the growing mode. In these plots, the modes with purely
real wavenumber are shown in black, the growing modes with $\Im\left(k\right)>0$
are shown in red and the decaying modes with $\Im\left(k\right)<0$
are shown in blue.\label{fig: ExampleDispersion}}
\end{figure}

In Examples (1)-(3) provided in Subsections \ref{exa:Example1}-\ref{exa:Example3},
we examine our developed method where the electron stream parameters
are $\check{\sigma}_{\mathrm{B}}=5.524\times10^{-21}$, $R_{\mathrm{sc}}=1$,
$\check{\omega}_{\mathrm{p}}=1$, and $\breve{v}=1$. We use dimensionless
variables (with inverted hat sign) and the normalization factor for
each physical quantity is described in Appendix \ref{sec:AppendixDimensionless}.
We consider the equivalent transmission line parameters as $C=16.667\:\left(=1.854\:\mathrm{nF/m}\right)$
and $L=1.5\times10^{-2}\:\mathrm{s^{2}/cm^{2}}\:\left(=1.348\:\mathrm{TH/m}\right)$
and we assume that the electron stream and electromagnetic wave are
in strong coupling by considering $b=1$. Then, the normalized TWT
principal parameter is calculated as $\check{\gamma}=0.234$. In general,
the interaction between the electromagnetic wave and the charge wave
occurs when they are synchronized, i.e., when the cold electromagnetic
wave phase velocity $w=1/\sqrt{LC}$ and the average velocity of the
electrons $\mathring{v}$ are matched. In the specified frequency
range, the cold electromagnetic wave phase velocity $w$ is assumed
to be dispersionless due to the dispersionless per-unit-length series
inductance $L$ and shunt capacitance $C$. The modal dispersion diagram
of the hot modes in the presented TWT is shown in Figures \ref{fig: ExampleDispersion}(a)-(d)
by using the previously mentioned parameters for TWT and substituting
them in Equation (\ref{eq: DispersionShortDimensionless}).

Due to the centrosymmetry of the set of modes in the TWT system (with
respect to $\omega=0$), we only show positive frequencies. This is
because the modes of negative frequencies can be easily recovered
from the centrosymmetry transformation. Centrosymmetry is a property
where for every point in the plot, there exists another point directly
opposite to it through zero-frequency. In other words, if a point
is on the plot, the same point reflected through the center will also
be on the plot. The conventional dispersion relation of the hot modes
is defined as the relation between real-valued angular frequency $\omega$
and complex-valued wavenumber $k$. The dispersion plots integrate
the dispersion relations of TWT system modal branches and provide
partial information on TWT system instabilities. The complete information
on the TWT system dispersion relation is encoded in the characteristic
equations for the complex-valued phase velocity of the hot modes (see
Equation (\ref{eq: Dispersion})). The characteristic equation has
exactly four complex-valued $k$ solutions for every $\omega$, taking
into account their algebraic multiplicity. Moreover, every hot mode
in the TWT system is determined by its frequency and wavenumber, that
is, by the pair $\left(\omega,k\left(\omega\right)\right)$ with complex-valued
$k\left(\omega\right)$.

In order to represent the convective (i.e., in space) unstable and
stable (oscillatory) modes of the TWT system, we proceed as follows.
Our first step is to parameterize every mode in the TWT system uniquely
using the pair $\left(\omega,k\left(\omega\right)\right)$. If $k\left(\omega\right)$
is degenerate, it is counted a relevant number of times according
to its multiplicity. Because of the importance of mode instability,
that is when $\Im\left\{ k\left(\omega\right)\right\} \neq0$, we
divide all the TWT system modes represented by pairs $\left(\omega,k\left(\omega\right)\right)$
into two distinct classes of unstable and stable (oscillatory) modes,
based on the complex velocity of hot modes $u\left(\omega\right).$
When the wavenumber of the hot mode $k\left(\omega\right)=\omega/u\left(\omega\right)$
is purely real or complex-valued with negative imaginary part, we
can consider the associated mode as a stable (oscillatory) mode. Oppositely,
we refer to the TWT system mode as a convective unstable mode if the
wavenumber of hot mode $k\left(\omega\right)$ is complex-valued with
a positive imaginary part. Notice that every point $\left(\omega,\Re\left\{ k\left(\omega\right)\right\} \right)$
in the convective unstable mode is associated with two complex conjugate
TWT system modes with $\pm\Im\left\{ k\left(\omega\right)\right\} $,
which only one of them with positive value leads to amplification
and the another mode with negative value will decay. There are two
regions of operation in an interactive system based on four hot eigenmodes:
(i) Amplification region: in this region the four modes are divided
into two sets of modes: the first set consists of two exponentially
growing and decaying oscillatory modes (amplifying/decaying, $\Im\left(k\right)\neq0$)
such that two modes wavenumbers are complex conjugate to each other
(i.e., $k_{1}=k_{2}^{*}$); the second set consists of two convectively
stable (oscillatory) modes (unamplifying/undecaying, $\Im\left(k\right)=0$)
that vary harmonically in time and are bounded in the entire space
by a constant; (ii) Non-amplified region: in this region the four
modes are convectively stable (oscillatory) with real-valued wavenumbers
(i.e., $\Im\left(k\right)=0$).

We define a critical transition point separating stability from instability
in the modal dispersion diagram. A transition point is a point $\left(\omega_{\mathrm{c}},k_{\mathrm{c}}\right)$
in the $\omega-k$ plane that marks a transitional point from stability
(oscillation) to instability (exponential growth) region. Transition
points are the points at which dispersion relations develop second-order
degeneracy. We refer to $\omega_{\mathrm{c}}$ as nodal frequencies
(transition frequencies) and $k_{\mathrm{c}}$ as nodal wavenumbers
(transition wavenumbers), which are indicated by green dots in Figure
\ref{fig: ExampleDispersion}(a)-(d). The amplification regime starts
or stops at this frequencies, and a deeper investigation into the
features of these critical points (they are exceptional points of
degeneracy) is provided in \cite{rouhi2021exceptional,figotin2021exceptional}.
The complex velocity changes dramatically at this critical point,
which is one of the distinctive features of the transition point (or
exceptional point). The real and imaginary parts of the complex velocity
of the hot mode $u=\omega/k$ with approximately frequency-independent
value at the high frequency are shown in Figure \ref{fig: ExampleDispersion}(e)-(f).
Also, the power gain factor $\alpha_{0}$ is illustrated in Figure
\ref{fig: ExampleDispersion}(g). In these plots, the black curves
show the modes with purely real wavenumber, the red curves show the
growing modes with $\Im\left(k\right)>0$ and the blue curves show
the decaying modes with $\Im\left(k\right)<0$.

In particular, in Subsection \ref{subsec:Exact-Method}, we investigated
pulse propagation along the TWT by using the \textquotedbl exact
method\textquotedblright{} and we will employ this method in Examples
(1)-(4) below. In addition, in Subsection \ref{subsec:Approximate-method},
we studied pulse propagation along the TWT by using the \textquotedbl approximate
method\textquotedblright{} (under constant $u$ assumption) and we
will apply this method in Examples (1) and (2) as a benchmark. We
also compare the results obtained by the exact and approximate method
in Examples (1) and (2) and discuss when the approximate method provides
reasonable accuracy.

\subsection{Example (1) - Wavepacket frequency band is far from the transition
points\label{exa:Example1}}

\begin{figure}
\centering{}\includegraphics[width=1\textwidth]{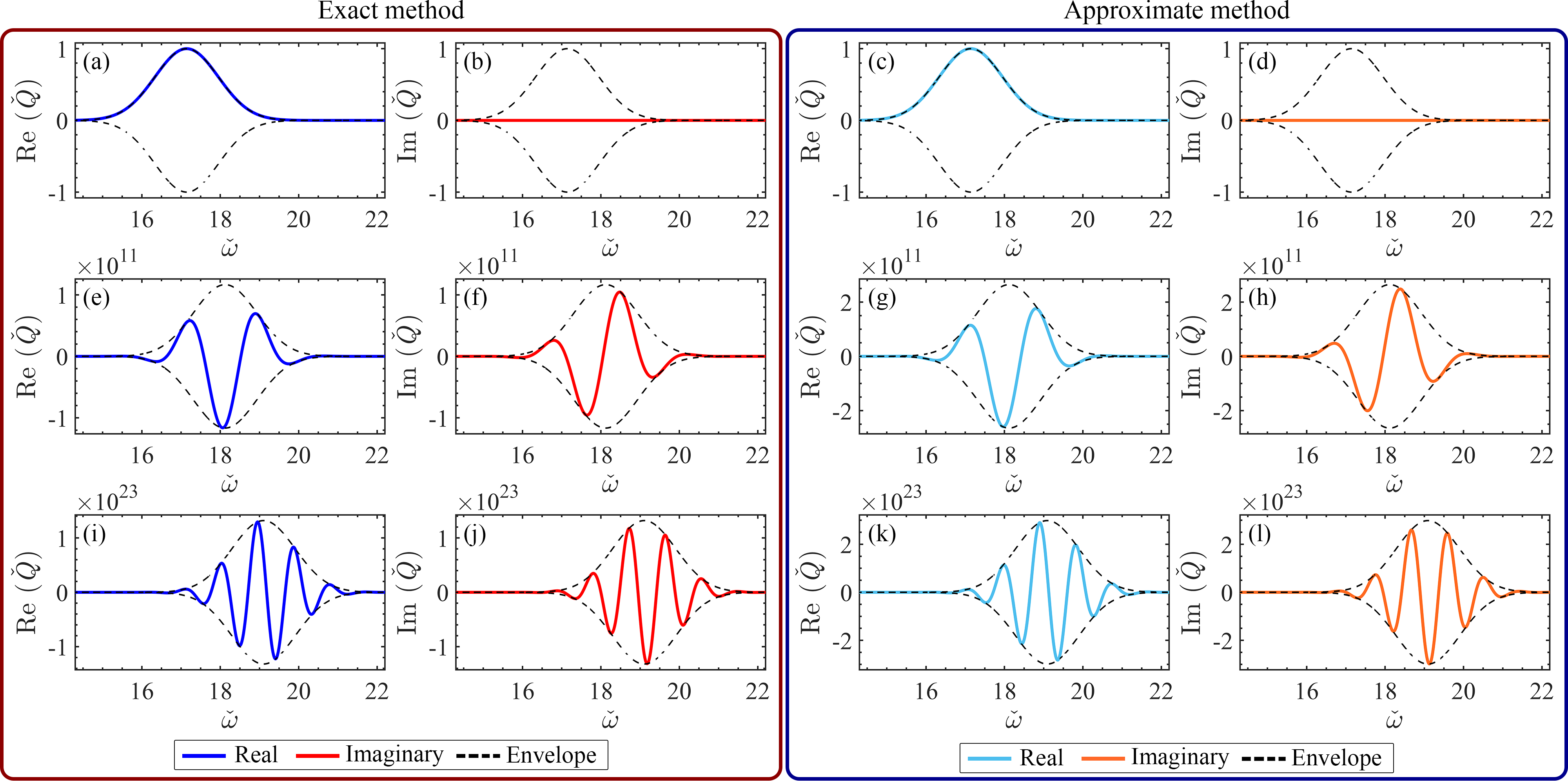}
\caption{Pulse propagation when the operating frequency is \textit{far} from
the transition points frequency. Plots (a)-(l) display the propagating
wavepackets of Example (1) in Subsection \ref{exa:Example1} at different
$z-$positions inside the TWT. These plots show the real (blue/cyan)
and imaginary (red/orange) parts at: (1) $\breve{z}=0$ - first row
from the top; (2) $\breve{z}=\breve{z}_{0}$ - second row from the
top; (3) $\breve{z}=2\breve{z}_{0}$ - third row from the top. The
first and second columns show the real and imaginary parts of the
propagated wavepacket calculated using the exact method. The third
and fourth columns display the real and imaginary parts of the propagated
wavepacket calculated under the constant $u$ assumption. We observe
amplification, shift in center frequency, and increase in the number
of local peaks in the wavepacket spectrum within the pulse bandwidth
as the pulse propagates through TWT.\label{fig: Example1Result}}
\end{figure}

In this example, we study wavepacket propagation when the operating
frequency and the pulse bandwidth are far from the transition points.
We show how under this condition the approximate method can calculate
the propagated wavepacket with acceptable accuracy. The input pulse
parameters are set as $A=1\:\mathrm{Fr}$, $\breve{\tau}_{\mathrm{w}}=0.279$
and $\check{\omega}_{0}=17.143$ for this example. We use dimensionless
variables, and the normalization factor for each quantity is described
in Appendix \ref{sec:AppendixDimensionless}; for this example the
length, time and frequency normalization factors are $\lambda_{\mathrm{p}}=2.691\times10^{10}\:\mathrm{cm}$,
$T_{\mathrm{p}}=0.898\:\mathrm{s}$ and $\omega_{\mathrm{p}}=7\:\mathrm{rad/s}$,
respectively. Based on the selected data, we calculate the propagated
wavepacket in the SWS at certain positions using the method described
in Subsection \ref{subsec:Fundamental-equations}. Then we compare
the exact calculated wavepacket with the results obtained under the
constant $u$ assumption. We start from the input port and show the
results for the propagated wavepacket at different positions in Figure
\ref{fig: Example1Result}. The solid blue curves show the real part,
the solid red curves depict the imaginary part, and the black dashed
curves show the absolute value or envelope of the propagating wavepacket
at different positions. Then, we move along the TWT and increase $\breve{z}=0$
to $\breve{z}=\breve{z}_{0}$, where $\breve{z}_{0}=3\times10^{-11}$.
Calculated results in Figures \ref{fig: Example1Result}(e) and (f)
show that the center angular frequency is increased by $\check{\omega}_{0}\times\delta_{\mathrm{w}}^{\mathrm{ex}}\left(\breve{z}_{0}\right)=0.98$,
and according to Equation (\ref{eq: delta_w}) the angular frequency
shift is obtained as $\check{\omega}_{0}\times\delta_{\mathrm{w}}^{\mathrm{app}}\left(\breve{z}_{0}\right)=0.974$.
At this point, we obtain the amplification factor $S_{\mathrm{amp}}^{\mathrm{ex}}\left(z\right)$ as
a natural logarithm of the maximum amplitude of the propagated wavepacket
that is obtained at the shifted center frequency. Figure \ref{fig: Example1Result}(e)
shows that the amplification factor is $S_{\mathrm{amp}}^{\mathrm{ex}}\left(\breve{z}_{0}\right)=25.482$,
and theoretical approximation based on Equation (\ref{eq: S_amp})
demonstrates that $S_{\mathrm{amp}}^{\mathrm{app}}\left(\breve{z}_{0}\right)=26.3$.
In order to compare the results extracted from exact method and the
approximate method under the constant $u$ assumption, the real and
imaginary parts of the propagated wavepacket at $\breve{z}=\breve{z}_{0}$
under the constant $u$ assumption are calculated and shown in Figures
\ref{fig: Example1Result}(g) and (h).

Next, we increase $\breve{z}$ to $2\breve{z}_{0}$ to analyze the
propagated pulse at the farther point. The propagated wavepacket obtained
by using the exact method are represented in Figures \ref{fig: Example1Result}(i)
and (j). We find that the center frequency has been increased and
the amplitude of the wavepacket has been significantly boosted. The
calculated result demonstrates that the center angular frequency is
increased by $\check{\omega}_{0}\delta_{\mathrm{w}}^{\mathrm{ex}}\left(2\breve{z}_{0}\right)=1.96$
and theoretical approximation under constant $u$ assumption predicts
$\check{\omega}_{0}\delta_{\mathrm{w}}^{\mathrm{app}}\left(2\breve{z}_{0}\right)=1.948$.
Also, we found that the amplification factor at this point is $S_{\mathrm{amp}}^{\mathrm{ex}}\left(2\breve{z}_{0}\right)=53.237$,
and approximate calculations result in $S_{\mathrm{amp}}^{\mathrm{app}}\left(2\breve{z}_{0}\right)=54.053$.
The number of local peaks in the frequency domain increases at further
distances from the input, compared to the results obtained for the
points near the input. In order to evaluate the error in comparison
of exact and approximate methods, the error function is defined as

\begin{equation}
\mathcal{E}r\left(X\right)=\left|\frac{X^{\mathrm{app}}-X^{\mathrm{ex}}}{X^{\mathrm{ex}}}\right|,
\end{equation}
where $X$ can be $\omega_{\mathrm{w}}\left(z\right)$ or $S_{\mathrm{amp}}\left(z\right)$,
and superscript \textquotedblleft app\textquotedblright{} shows the
approximate value under constant $u$ assumption and superscript \textquotedblleft ex\textquotedblright{}
indicates the exact value based on analytical exact method. Finally,
the calculated results are summarized in Table \ref{tab: Example1Compare}.
It demonstrates that the exact calculation and the prediction under
the assumption of a constant $u$ are in very close agreement. This
is because the center frequency of the input pulse is chosen in the
dispersionless region with approximately constant hot mode complex
velocity $\overline{u}$.

\begin{table}
\begin{centering}
\caption{Comparison between calculated results for Gaussian pulse propagation
under the constant $u$ assumption (indicated by app) versus the exact
results obtained by using Equation (\ref{eq: Differential Equation})
(indicated by ex) in the Example (1) in Subsection \ref{exa:Example1}.\label{tab: Example1Compare}}
\par\end{centering}
\centering{}%
\begin{tabular}{|c|c|c|c|c|c|c|}
\hline 
 & $\check{\omega}_{\mathrm{w}}^{\mathrm{ex}}\left(z\right)$ & $\check{\omega}_{\mathrm{w}}^{\mathrm{app}}\left(z\right)$ & $S_{\mathrm{amp}}^{\mathrm{ex}}\left(z\right)$ & $S_{\mathrm{amp}}^{\mathrm{app}}\left(z\right)$ & $\mathcal{E}r\left(\omega_{\mathrm{w}}\left(z\right)\right)$ & $\mathcal{E}r\left(S_{\mathrm{amp}}\left(z\right)\right)$\tabularnewline
\hline 
\hline 
$\breve{z}=0$ & $17.143$ & $17.143$ & $0$ & $0$ & $0\%$ & $0\%$\tabularnewline
\hline 
$\breve{z}=3\times10^{-11}$ & $18.123$ & $18.117$ & $25.482$ & $26.3$ & $0.034\%$ & $3.211\%$\tabularnewline
\hline 
$\breve{z}=6\times10^{-11}$ & $19.103$ & $19.091$ & $53.237$ & $54.053$ & $0.06\%$ & $1.533\%$\tabularnewline
\hline 
\end{tabular}
\end{table}

\subsection{Example (2) - Wavepacket frequency band is near the transition points\label{exa:Example2}}

\begin{figure}
\centering{}\includegraphics[width=1\textwidth]{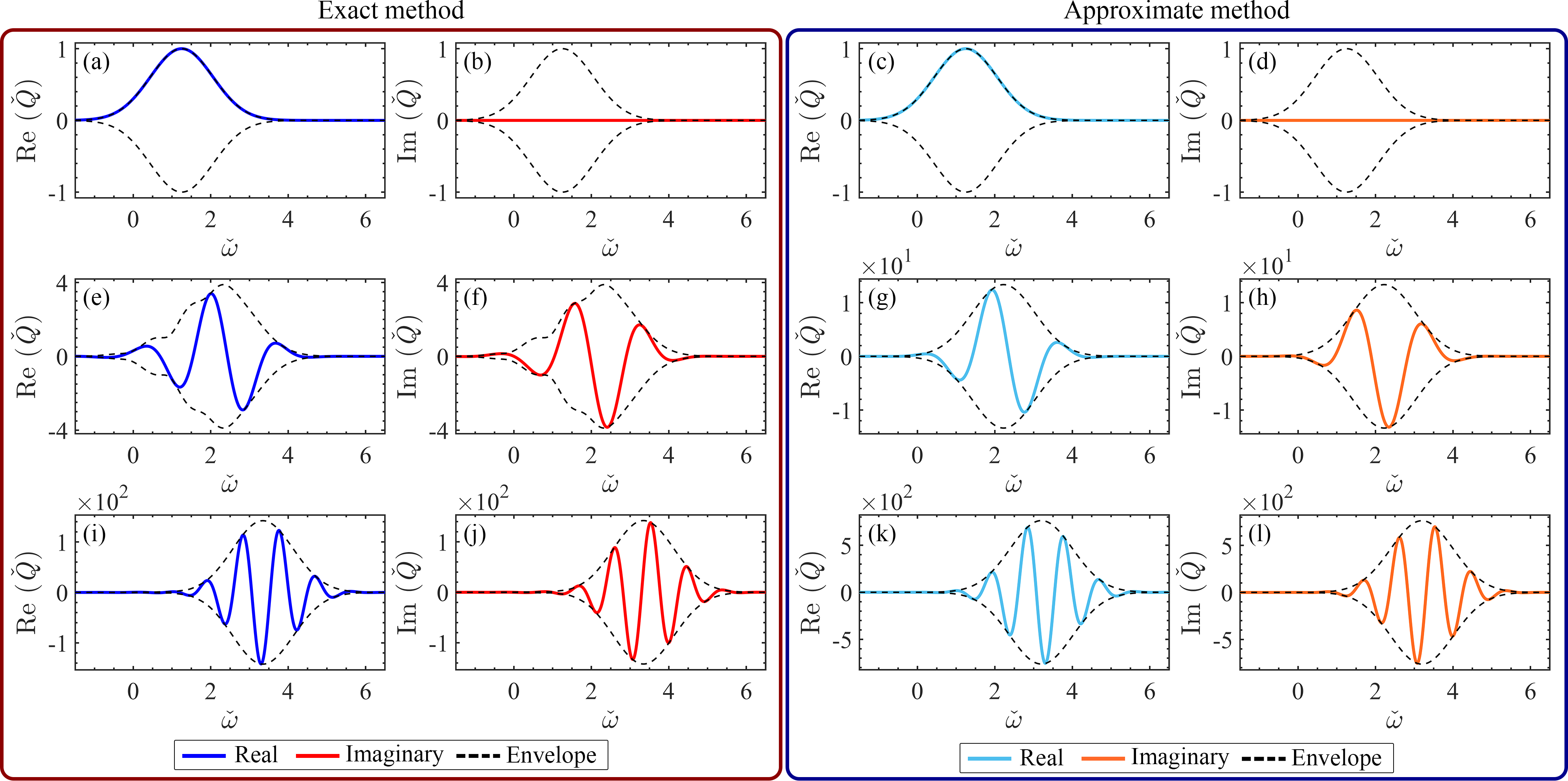}
\caption{As in Figure \ref{fig: Example1Result}, but for the case when the
pulse bandwidth \textit{contains} the transition point. Pulses in
the second row are significantly distorted compared to those in the
first row due to the pulse spectrum proximity to the third transition
point (plots (e) and (f)). The approximate results under the constant
$u$ assumption (plots (g) and (h)) cannot predict the propagated
pulse shape and amplitude correctly.\label{fig: Example2Result}}
\end{figure}

In Example (1) in Subsection \ref{exa:Example1}, we assumed that
the wavepacket center frequency was far enough from the transition
points. Consequently, the constant value assumption for hot mode complex
velocity was valid in our computations. \textit{But the question is,
does this assumption work in other cases as well?} In order to answer
the question, we study another example where the pulse bandwidth \textit{contains}
the transition frequency. Let us assume $\check{\omega}_{0}=\check{\omega}_{\mathrm{c},3}=1.25$
where $\check{\omega}_{\mathrm{c},3}$ is the frequency of the third
transition point, as shown in Figure \ref{fig: ExampleDispersion}.
The calculated propagated wavepacket by using the mentioned pulse
center frequency are illustrated in Figure \ref{fig: Example2Result},
and significant outcomes are summarized in Table \ref{tab: Example2Compare}.
The calculated results in Figure \ref{fig: Example2Result} show that
both center frequency and field amplitude exponent are increased by
moving throughout the TWT in a forward direction. In addition, the
number of local peaks in the wavepacket spectrum within the pulse
bandwidth is increased gradually. The real and imaginary parts of
the propagated wavepacket using an approximate method are shown in
the third and fourth columns of Figure \ref{fig: Example2Result}.
According to the calculated error in Table \ref{tab: Example2Compare},
we observe a significant deviation in the approximate results compared
to the exact calculation. The presented example demonstrates that
the approximate method under constant $u$ assumption cannot deliver
precise results if the center frequency is near transition points
or if the pulse bandwidth contains a transition point. Thus, we must
be sure that $\check{\omega}_{0}>>\check{\omega}_{\mathrm{c}}$ to
use the advantages of approximate method for calculating the propagated
wavepacket. However, the approximate method was only used as a benchmark
and the developed exact method is efficient and fast enough to use
in practical applications.

\begin{table}
\caption{Comparison between calculated results for Gaussian pulse propagation
under the constant $u$ assumption (indicated by app) versus the exact
results obtained by using Equation (\ref{eq: Differential Equation})
(indicated by ex) in the Example (2) in Subsection \ref{exa:Example2}.\label{tab: Example2Compare}}

\centering{}%
\begin{tabular}{|c|c|c|c|c|c|c|}
\hline 
 & $\check{\omega}_{\mathrm{w}}^{\mathrm{ex}}\left(z\right)$ & $\check{\omega}_{\mathrm{w}}^{\mathrm{app}}\left(z\right)$ & $S_{\mathrm{amp}}^{\mathrm{ex}}\left(z\right)$ & $S_{\mathrm{amp}}^{\mathrm{app}}\left(z\right)$ & $\mathcal{E}r\left(\omega_{\mathrm{w}}\left(z\right)\right)$ & $\mathcal{E}r\left(S_{\mathrm{amp}}\left(z\right)\right)$\tabularnewline
\hline 
\hline 
$\breve{z}=0$ & $1.25$ & $1.25$ & $0$ & $0$ & $0\%$ & $0\%$\tabularnewline
\hline 
$\breve{z}=3\times10^{-11}$ & $2.327$ & $2.224$ & $1.357$ & $2.591$ & $4.436\%$ & $90.922\%$\tabularnewline
\hline 
$\breve{z}=6\times10^{-11}$ & $3.346$ & $3.198$ & $4.957$ & $6.636$ & $4.416\%$ & $33.876\%$\tabularnewline
\hline 
\end{tabular}
\end{table}

There is distortion in the envelope of the propagated wavepacket in
the second row of Figure \ref{fig: Example2Result} due to nontrivial
phenomena of non-Gaussian pulses in the system. The exact result is
distorted at lower frequencies, but the approximate result is not.
This is due to the proximity of the wavepacket bandwidth to the transition
point. This phenomenon can be better understood by looking at the
following example.

\subsection{Example (3) - Variation in the center frequency of the wavepacket\label{exa:Example3}}

\begin{figure}
\centering{}\includegraphics[width=0.75\textwidth]{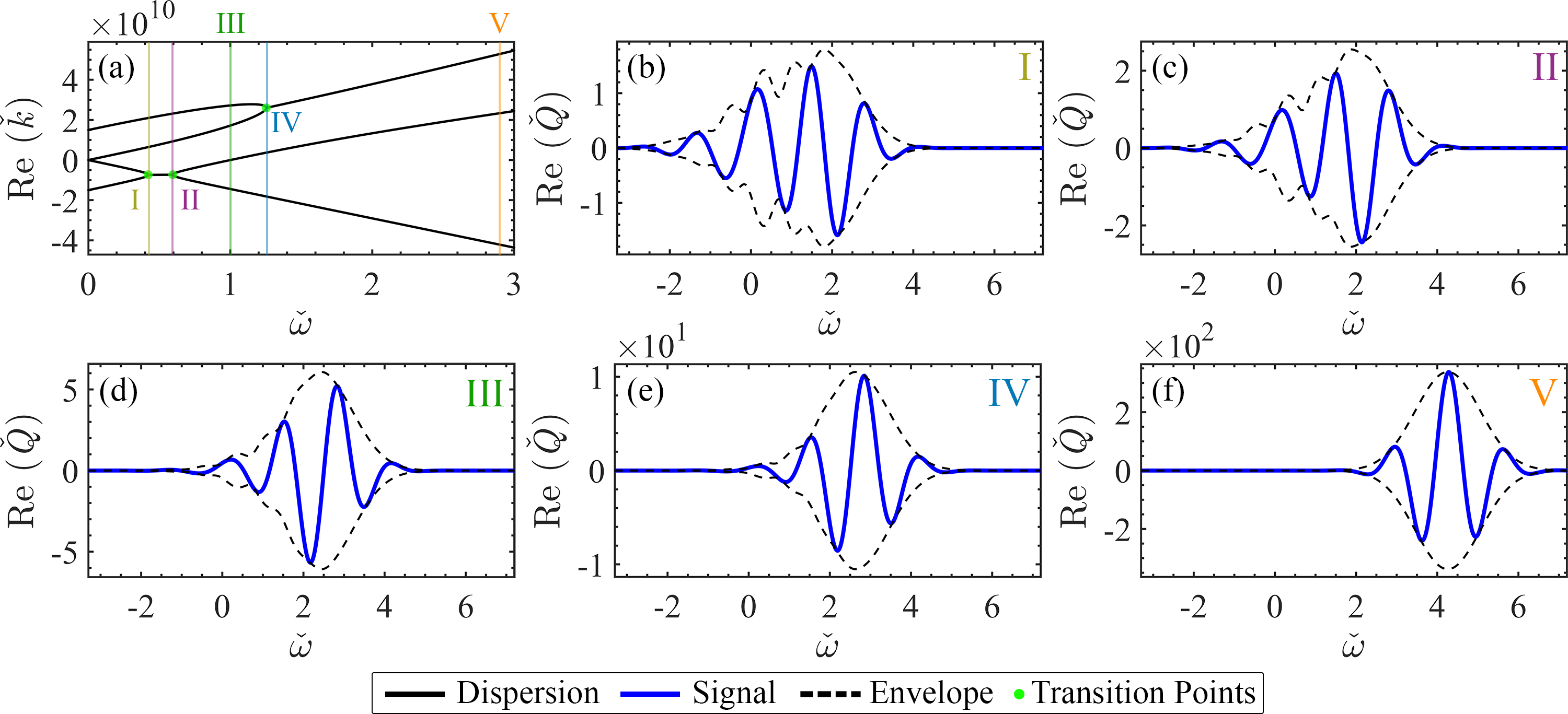}
\caption{Pulse propagation in different frequency regimes. Plot (a) shows the
real part of the hot mode eigenvalues in the Example (3) in Subsection
\ref{exa:Example3}, and it also indicates the selected center frequencies
of the input pulse for five cases. Plots (b)-(f) display the calculated
results for distinct center frequencies of the input Gaussian pulse,
namely the real parts of pulse for an observer located at $\breve{z}=4\times10^{-11}$
in the various center frequency of (b) $\check{\omega}_{0}=0.423$,
(c) $\check{\omega}_{0}=0.594$, (d) $\check{\omega}_{0}=1$, (e)
$\check{\omega}_{0}=1.253$, and (f) $\check{\omega}_{0}=2.9$. The
propagated wavepacket is distorted when the center frequency of the
input pulse is near a transition point (Cases I, II, III and IV).
The propagated pulse completely maintains its Gaussian shape when
the center frequency of the input pulse is far enough away from the
transition point (dispersionless region) since the hot mode complex
velocity is constant (Case V).\label{fig: CenterFrequencyVariation}}
\end{figure}

The purpose of this example is to examine the effect of changing the
center frequency of the input pulse on the shape of the pulse envelope
after it travels along the TWT. As we discussed in Examples (1) and
(2) in Subsections \ref{exa:Example1} and \ref{exa:Example2}, we
should select the proper operating frequency to avoid distortion in
the wavepacket. Let us consider a TWT with the parameters used in
previous examples and the same parameters for the Gaussian input pulse.
Then, we vary the center frequency of the wavepacket and calculate
the propagated wavepacket at $\breve{z}=4\times10^{-11}$. We select
five different center frequencies for the input wavepacket, indicated
by different colors in the dispersion diagram in Figure \ref{fig: CenterFrequencyVariation}(a).
The first input frequency is chosen at the first transition point,
where the center angular frequency is $\check{\omega}_{0}=\check{\omega}_{\mathrm{c},1}=0.423$.
Then the second input frequency is chosen at the second transition
point, which is $\check{\omega}_{0}=\check{\omega}_{\mathrm{c},2}=0.594$.
The results for these two center frequencies at $\breve{z}=4\times10^{-11}$
are shown in Figures \ref{fig: CenterFrequencyVariation}(b) and (c).
The propagated wavepacket exhibits significant distortion since the
pulse bandwidth contains the transition point. Next, we increase the
center angular frequency to $\check{\omega}_{0}=1$ and $\check{\omega}_{0}=\check{\omega}_{\mathrm{c},3}=1.253$.
The calculated results for these two cases are shown in Figures \ref{fig: CenterFrequencyVariation}(d)
and (e), where a small distortion in the shape of the propagated wavepacket
can be recognized. In the case shown in Figure \ref{fig: CenterFrequencyVariation}(e),
the lower frequencies of the pulse bandwidth are distorted since it
contains a transition point. In contrast, the higher frequencies of
the pulse bandwidth are amplified without any distortion since they
are located in the dispersionless region with a constant value of
hot mode complex velocity. Finally, we select the center frequency
of the input pulse in the dispersionless region, which is far from
the transition points. In this case, $\check{\omega}_{0}=2.9$ and
the input pulse is amplified without distortion as illustrated in
Figure \ref{fig: CenterFrequencyVariation}(f).

\subsection{Example (4) - Realistic Helix TWT\label{exa:ExampleHelix}}

\begin{figure}
\centering{}\includegraphics[width=0.68\textwidth]{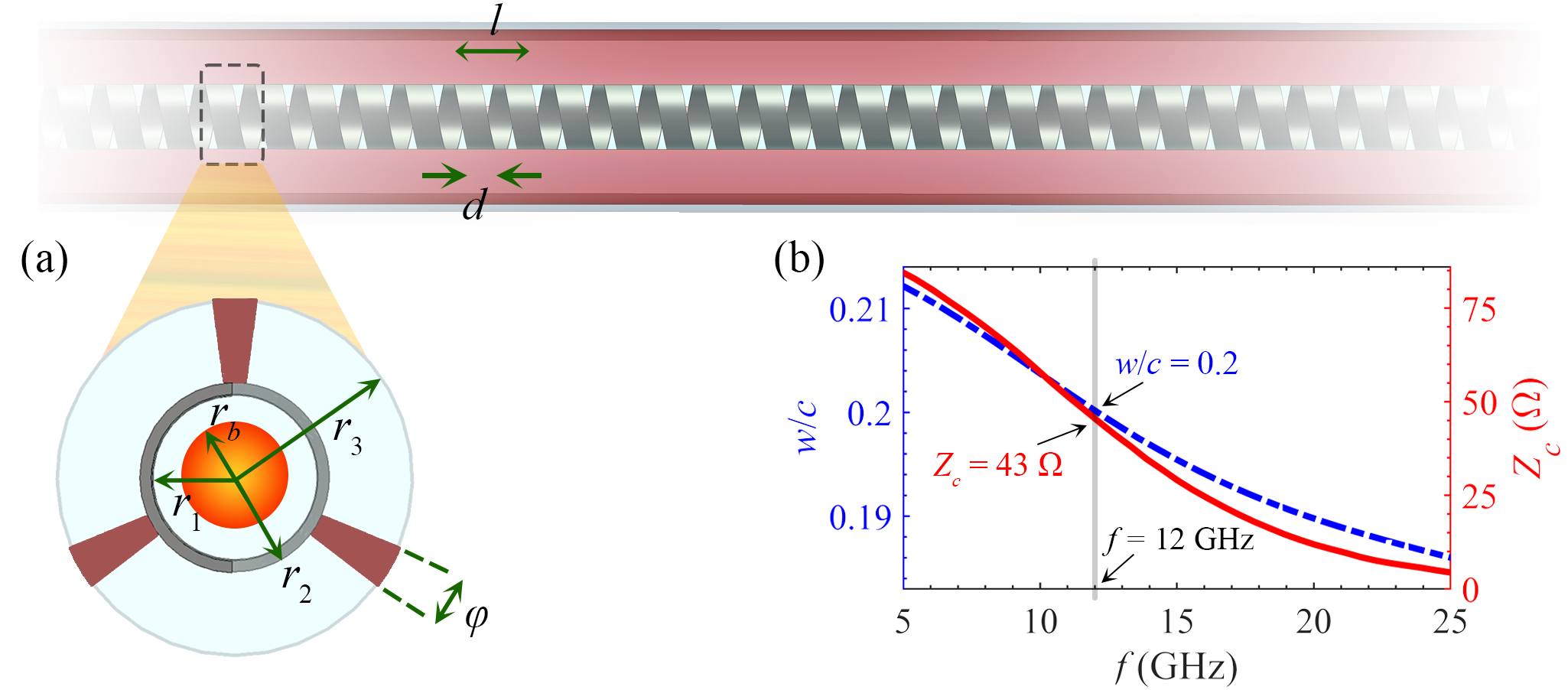}
\caption{(a) Schematic of the tape helix SWS in a circular metallic waveguide.
(b) Cold electromagnetic wave phase velocity and characteristic impedance
of the first forward mode of the helix SWS obtained from cold full-wave
simulations in the absence of the electron stream, using the finite-element
method eigenmode solver.\label{fig: RealisticScheme}}
\end{figure}

In order to evaluate the presented analysis for a specific design
in the microwave regime, we use the characteristic parameters of a
helix TWT with realistic settings in this example. The helix TWT is
illustrated in Figure \ref{fig: RealisticScheme}(a), which was already
designed and utilized in \cite{rouhi2021exceptional} and \cite{abdelshafy2022accurate}.
In a helix-based SWS, a metallic tape-helix is surrounded by a metallic
waveguide \cite{watkins1954helix}. The helix SWS utilizes a conventional
two-body cylindrical vacuum envelope that contains a metallic tape
helix supported by three dielectric BeO rods \cite{han2008thermal}.
An electron stream flows along the axis of the helical conductor with
inner radius $r_{1}=7.44\times10^{-2}\:\mathrm{cm}$, and outer radius
$r_{2}=8.46\times10^{-2}\:\mathrm{cm}$. The metallic circular waveguide
has a radius of $r_{3}=1.06\times10^{-1}\:\mathrm{cm}$ and the three
equally spaced dielectric rods support that physically hold the helix
are made of BeO with a relative dielectric constant of $\varepsilon_{\mathrm{r}}=6.5$.
Moreover, the other geometric parameters are $l=1.04\times10^{-1}\:\mathrm{cm}$,
$d=5.2\times10^{-2}\:\mathrm{cm}$, and $\varphi=14.2\lyxmathsym{\textdegree}$.
The input and output RF pulses of the structure are defined as RF
input port and RF output port as shown in Figure \ref{fig: PulseProp}.
A finite-element eigenmode solver in CST Studio Suite by DS SIMULIA
is used to simulate the helix SWS. The eigenmode solver enforces a
phase shift across the structure period in the longitudinal direction
of propagation and solves for the real-valued eigenfrequencies in
the absence of electron stream. The simulation is repeated for each
phase shift to extract the characteristic parameters, i.e., the cold
electromagnetic wave phase velocity and the equivalent transmission
line characteristic impedance. Then, the calculated parameters for
the mentioned geometry are illustrated in Figure \ref{fig: RealisticScheme}(b).
For the sake of simplicity, we assume constant values for the cold
electromagnetic wave phase velocity and characteristic impedance.
In this example, we use values around the synchronization point, which
is $f_{\mathrm{sync}}=12\:\mathrm{GHz}$. The maximum interaction
between the electromagnetic and the space-charge wave occurs when
they are synchronized, i.e., by matching $w$ and $\mathring{v}$.
The selected values for this example are chosen as $w/c=0.2c$ and
$Z_{\mathrm{c}}=43\:\Omega$. By using the extracted values for $w$
and $Z_{\mathrm{c}}$ at the synchronization point, the equivalent
distributed series inductance $L=Z_{\mathrm{c}}/w$ and shunt capacitance
$C=1/\left(Z_{\mathrm{c}}w\right)$ are calculated. For this case,
we obtain $L=717.162\:\mathrm{nH/m}\:\left(=7.980\times10^{-21}\,\mathrm{\mathrm{s^{2}}/cm^{2}}\right)$
and $C=387.865\:\mathrm{pF/m}\:\left(=3.486\right)$. Such TWT amplifier
uses a solid linear electron stream with a radius of $r_{\mathrm{b}}=5.6\times10^{-2}\:\mathrm{cm}$.
The electron stream phase velocity is $0.2$ times the speed of light
($\mathring{v}=0.2c$) to have synchronization around $f_{\mathrm{sync}}=12\:\mathrm{GHz}$.
The value of the emitted current was set to $I_{0}=47\:\mathrm{mA\:(=1.41\times10^{8}\:\mathrm{StatA})}$
and the corresponding plasma angular frequency is $\omega_{\mathrm{p}}=2\pi\times624.6\times10^{6}\:\mathrm{rad/s}$.
We use a plasma frequency reduction factor of $R_{\mathrm{sc}}=0.12$,
which was calculated and used in \cite{rouhi2021exceptional} and
\cite{abdelshafy2022accurate}. According to the calibration used
in \cite{rouhi2021exceptional}, the value of the coupling strength
factor $b$, which is an essential parameter in the Lagrangian model,
is estimated as $b=0.92$. By using the introduced parameters, we
calculate the required parameters of the Lagrangian model as $\beta=1.739\times10^{14}\:\mathrm{cm^{2}/s^{2}}$
and $\gamma=4.194\times10^{13}\:\mathrm{cm^{2}/s^{2}}$. Additionally,
the dimensionless version of parameters required for the dimensionless
form of the dispersion equation in Equation (\ref{eq: DispersionShortDimensionless})
are expressed as $\check{w}=1$ and $\check{\gamma}=1.167\times10^{-6}$.

\begin{figure}
\centering{}\includegraphics[width=1\textwidth]{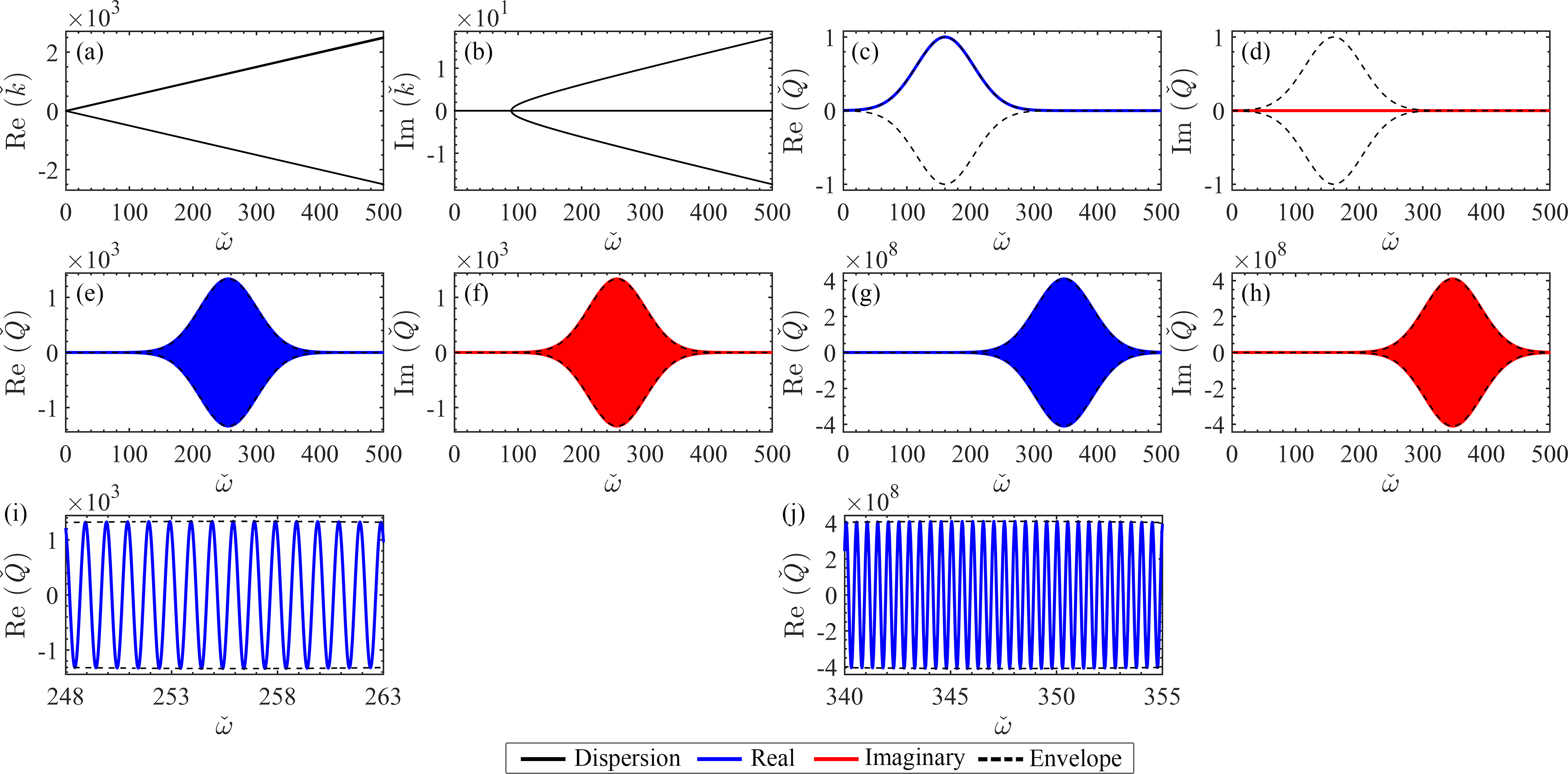}
\caption{(a) Real and (b) imaginary parts of the complex-valued wavenumber
for the Example (4) in Subsection \ref{fig: ExampleRealistic}. Plots
(c)-(h) display the real and imaginary part of the spectrum of the
propagated wavepacket at different positions along the TWT at: (c)
and (d) $\breve{z}=0$; (e) and (f) $\breve{z}=0.2$; (g) and (h)
$\breve{z}=0.4$. The calculated results are obtained by using Equation
(\ref{eq: Differential Equation}). Plots (i) and (j) show the enlarged
version of the real parts of the spectrum of the propagated wavepacket
at $\breve{z}=0.2$ and $\breve{z}=0.4$ in the frequency range of
$\check{\omega}=248$ to $\check{\omega}=263$ ($\Delta\check{\omega}=15$)
and $\check{\omega}=340$ to $\check{\omega}=355$ ($\Delta\check{\omega}=15$).
These two plots demonstrate that the number of local peaks in the
wavepacket spectrum is increased in the same bandwidth by moving along
the TWT.\label{fig: ExampleRealistic}}
\end{figure}

Next, we study the nontrivial phenomena during the Gaussian pulse
propagation for this specific design. The real and imaginary parts
of the complex-valued wavenumber of hot modes are shown in Figures
\ref{fig: ExampleRealistic}(a) and (b). In the proposed example,
parameters of the RF input pulse are set as $A=10^{4}\:\mathrm{C}\:(=2.998\times10^{13}\:\mathrm{Fr})$,
$\breve{\tau}_{\mathrm{w}}=5\times10^{-3}$ and $\check{\omega}_{0}=160.103$.
Since the center frequency of the RF input signal is sufficiently
far from transition points, we can consider the constant value for
the hot mode complex velocity $u$. The normalization factors used
for length, time and frequency in this example are $\lambda_{\mathrm{p}}=4.000\:\mathrm{m}$,
$T_{\mathrm{p}}=13.342\:\mathrm{ns}$ and $\omega_{\mathrm{p}}=2\pi\times624.6\times10^{6}\:\mathrm{rad/s}$
respectively. Then, we calculate the propagated wavepacket in the
SWS at certain positions using the analytical method described in
Section \ref{subsec:Fundamental-equations}. We start from the initial
point, i.e., input port at $\breve{z}=0$, and show the extracted
calculated results in Figures \ref{fig: ExampleRealistic} (c) and
(d). The solid blue curves depict the real part, solid red curves
depict the imaginary part, and black dashed curves indicate the absolute
value or envelope of the RF input wavepacket. Then, we move further
in the TWT and increase $\breve{z}=0$ to $\breve{z}=0.2$. Calculated
results in Figures \ref{fig: ExampleRealistic}(e) and (f) show that
the center frequency is increased and the wavepacket is amplified
dramatically. The huge amplification in the wavepacket occurred due
to imaginary parts of the dispersion diagram in the operating frequencies.
Next, we move further in the SWS and calculate the propagated wavepacket
at $\breve{z}=0.4$. The real and imaginary part of propagated wavepacket
are presented in Figures \ref{fig: ExampleRealistic}(g) and (h).
By comparing the calculated results at two selected points we observe
that the center frequency of the pulse is increased by traveling in
the SWS. Moreover, the magnified version of the real part of the wavepacket
in the same frequency range in Figures \ref{fig: ExampleRealistic}(i)
and (j) shows that the number of local peaks in the wavepacket spectrum
is increasing. This example explains nontrivial phenomena that can
happen during pulse propagation in a structure with realistic parameters.
For simplicity of analysis, we considered only a single-stage TWT
in this example. However, by leveraging our previously developed models
for multi-stage designs presented in \cite{marosi2024small,rouhi2025small},
the proposed analysis can be readily extended to multi-stage configurations.

\section{Conclusions\label{sec:Conclusions}}

We have introduced an efficient method for wavepacket propagation
analysis in TWTs based on the Lagrangian field theory. The resulting
Euler-Lagrange equations are second order differential equations in
both time and space. While a Gaussian waveform is assumed for simplicity,
the algorithm is applicable to arbitrary input pulses. In the case
of wave packets with spectral contents located in a linear region
of the dispersion diagram, the analytical method can provide accurate
results.

Overall, the proposed method offers valuable insight into wavepacket
dynamics in TWTs. In particular, the method reveals phenomena such
as super-exponential amplification, center-frequency shift, and an
increased number of local peaks in the wavepacket spectrum as the
pulse propagates through the TWT. The methodology contributes to a
deeper understanding of beam-wave interactions in vacuum electronics
and provides a foundation for further analytical and computational
advancements.

\textbf{Data Availability:} The data that support the findings of
this study are available within the article.

\textbf{Acknowledgment:} This work was supported by the Air Force
Office of Scientific Research (AFOSR) Multidisciplinary University
Research Initiative (MURI) under Grant No. FA9550-20-1-0409 administered
through the University of New Mexico. The authors are thankful to
DS SIMULIA for providing CST Studio Suite that was instrumental in
this study.

\section*{Appendix A. CGS to SI Conversion\label{sec:AppendixCGSSI}}

Table \ref{tab: CGSSI-1} lists the relationships between quantities
in SI and CGS (Gaussian) units \cite{jackson1998classical}. For greater
numerical precision, all factors of 3 may be replaced with 2.99792458,
reflecting the exact value of the speed of light in vacuum.

\begin{table}[h]
\caption{Conversion table for given amounts of a physical quantity.\label{tab: CGSSI-1}}

\centering{}%
\begin{tabular}{|c|c|c|}
\hline 
Physical Quantity & SI & CGS (Gaussian)\tabularnewline
\hline 
\hline 
Length ($l$) & $1$ meter (m) & $10^{2}$ centimeters (cm)\tabularnewline
\hline 
Time ($t$) & $1$ second (s) & $1$ second (s)\tabularnewline
\hline 
Frequency ($f$) & $1$ hertz (Hz) & $1$ hertz (Hz)\tabularnewline
\hline 
Charge ($q$) & $1$ coulomb (C) & $c_{\mathrm{G}}\times10^{-1}$ statcoulombs (FR)\tabularnewline
\hline 
Current ($I$) & $1$ ampere (A) & $c_{\mathrm{G}}\times10^{-1}$ statamperes (FR/s)\tabularnewline
\hline 
Voltage ($V$) & $1$ volt (V) & $c_{\mathrm{G}}^{-1}\times10^{8}$ statvolt (statV)\tabularnewline
\hline 
Capacitance ($C$) & $1$ farad (F) & $c_{\mathrm{G}}^{2}\times10^{-9}$ cm\tabularnewline
\hline 
Inductance ($L$) & $1$ henry (H) & $c_{\mathrm{G}}^{-2}\times10^{9}$ $\mathrm{s^{2}/cm}$\tabularnewline
\hline 
\multicolumn{3}{|c|}{$c_{\mathrm{SI}}=3\times10^{8}\:\mathrm{m/s}$}\tabularnewline
\multicolumn{3}{|c|}{$c_{\mathrm{G}}=3\times10^{10}\:\mathrm{cm/s}$}\tabularnewline
\hline 
\end{tabular}
\end{table}

Moreover, parameter conversions between the two unit systems---as
well as the corresponding equation transformations---are summarized
in Table \ref{tab: CGSSI-2}.

\begin{table}[h]
\caption{Conversion table for symbols and formulas.\label{tab: CGSSI-2}}

\begin{centering}
\begin{tabular}{|c|c|c|c|c|}
\hline 
Quantity & Symbol & SI unit & Gaussian unit & Conversion factor\tabularnewline
\hline 
\hline 
Electric charge & $q$ & $\mathrm{C}$ & $\mathrm{Fr}\:\left(\mathrm{cm}^{3/2}\mathrm{g}^{1/2}\mathrm{s}^{-1}\right)$ & $\frac{q_{\mathrm{G}}}{q_{\mathrm{SI}}}=\frac{1}{\sqrt{4\pi\varepsilon_{0}}}=\frac{3\times10^{9}\:\mathrm{Fr}}{1\:\mathrm{C}}$\tabularnewline
\hline 
Electric current & $I$ & $\mathrm{A}$ & $\mathrm{Fr/s}\:\left(\mathrm{cm}^{3/2}\mathrm{g}^{1/2}\mathrm{s}^{-2}\right)$ & $\frac{I_{\mathrm{G}}}{I_{\mathrm{SI}}}=\frac{1}{\sqrt{4\pi\varepsilon_{0}}}=\frac{3\times10^{9}\:\mathrm{Fr/s}}{1\:\mathrm{A}}$\tabularnewline
\hline 
Electric voltage & $V$ & $\mathrm{V}$ & $\mathrm{StatV}\:\left(\mathrm{cm}^{3/2}\mathrm{g}^{1/2}\mathrm{s}^{-1}\right)$ & $\frac{V_{\mathrm{G}}}{V_{\mathrm{SI}}}=\sqrt{4\pi\varepsilon_{0}}=\frac{1\:\mathrm{StatV}}{3\times10^{2}\:\mathrm{V}}$\tabularnewline
\hline 
Electric field & $\mathbf{E}$ & $\mathrm{V/m}$ & $\mathrm{StatV/cm}\:\left(\mathrm{cm}^{-1/2}\mathrm{g}^{1/2}\mathrm{s}^{-1}\right)$ & $\frac{\mathbf{E}_{\mathrm{G}}}{\mathbf{E}_{\mathrm{SI}}}=\sqrt{4\pi\varepsilon_{0}}=\frac{1\:\mathrm{StatV/cm}}{3\times10^{4}\:\mathrm{V/m}}$\tabularnewline
\hline 
Magnetic field & $\mathbf{H}$ & $\mathrm{A/m}$ & $\mathrm{Oe}\:\left(\mathrm{cm}^{-1/2}\mathrm{g}^{1/2}\mathrm{s}^{-1}\right)$ & $\frac{\mathbf{H}_{\mathrm{G}}}{\mathbf{H}_{\mathrm{SI}}}=\sqrt{4\pi\varepsilon_{0}}=\frac{4\pi\times10^{-3}\:\mathrm{Oe}}{1\:\mathrm{A/m}}$\tabularnewline
\hline 
Resistance & $R$ & $\mathrm{\Omega}$ & $\mathrm{s/cm}$ & $\frac{R_{\mathrm{G}}}{R_{\mathrm{SI}}}=4\pi\varepsilon_{0}=\frac{1\:\mathrm{s/cm}}{3^{2}\times10^{11}\:\mathrm{\Omega}}$\tabularnewline
\hline 
Capacitance & $C$ & $\mathrm{F}$ & $\mathrm{cm}$ & $\frac{C_{\mathrm{G}}}{C_{\mathrm{SI}}}=\frac{1}{4\pi\varepsilon_{0}}=\frac{3^{2}\times10^{11}\:\mathrm{cm}}{1\:\mathrm{F}}$\tabularnewline
\hline 
Inductance & $L$ & $\mathrm{H}$ & $\mathrm{s^{2}/cm}$ & $\frac{L_{\mathrm{G}}}{L_{\mathrm{SI}}}=4\pi\varepsilon_{0}=\frac{1\:\mathrm{s^{2}/cm}}{3^{2}\times10^{11}\:\mathrm{H}}$\tabularnewline
\hline 
Capacitance (p.u.l) & $\overline{C}$ & $\mathrm{F/m}$ & $\mathrm{dim-less}$ & $\frac{\overline{C}_{\mathrm{G}}}{\overline{C}_{\mathrm{SI}}}=\frac{1}{4\pi\varepsilon_{0}}=\frac{3^{2}\times10^{9}}{1\:\mathrm{F/m}}$\tabularnewline
\hline 
Inductance (p.u.l) & $\overline{L}$ & $\mathrm{H/m}$ & $\mathrm{s^{2}/cm^{2}}$ & $\frac{\overline{L}_{\mathrm{G}}}{\overline{L}_{\mathrm{SI}}}=4\pi\varepsilon_{0}=\frac{1\:\mathrm{s^{2}/cm^{2}}}{3^{2}\times10^{13}\:\mathrm{H/m}}$\tabularnewline
\hline 
\multicolumn{5}{|c|}{$\varepsilon_{0}=8.8541878\times10^{-12}\:\mathrm{F/m}$}\tabularnewline
\multicolumn{5}{|c|}{$\mu_{0}=1.2566370\times10^{-6}\:\mathrm{H/m}$}\tabularnewline
\hline 
\end{tabular}
\par\end{centering}
{*} p.u.l: per unit length

{*} dim-less: dimensionless
\end{table}

\section*{Appendix B. Pierce Parameters\label{sec:AppendixPierce}}

For convenience, Table \ref{tab: Pierce} presents the transformations
between the Lagrangian model parameters and the Pierce model parameters
used in this paper.

\begin{table}[h]
\caption{Translation from Lagrangian model parameters to the Pierce model parameters.\label{tab: Pierce}}

\centering{}%
\begin{tabular}{|c|c|c|}
\hline 
Parameter & Lagrangian model & Pierce model\tabularnewline
\hline 
\hline 
electron stream velocity & $\mathring{v}$ & $u_{0}$\tabularnewline
\hline 
electron stream cross-section area & $\sigma_{\mathrm{B}}$ & $A$\tabularnewline
\hline 
Stream intensity & $\beta$ & $\frac{I_{0}u_{0}}{2V_{0}}$\tabularnewline
\hline 
Characteristic phase velocity of cold electromagnetic modes & $w$ & $\sqrt{-\frac{\omega^{2}}{ZY}}$\tabularnewline
\hline 
TWT principal parameter & $\gamma$ & $j\omega\frac{a^{2}}{Y}\frac{I_{0}}{2V_{0}}u_{0}$\tabularnewline
\hline 
\end{tabular}
\end{table}

\section*{Appendix C. Dimensionless Variables\label{sec:AppendixDimensionless}}

This paper employs dimensionless variables, with the corresponding
normalization factors for each parameter provided in Table \ref{tab: Dimensionless}.

\begin{table}[H]
\caption{List of dimensionless variables with the corresponding normalization
factor.\label{tab: Dimensionless}}

\centering{}%
\begin{tabular}{|c|c|c|}
\hline 
Dimension & Normalization factor & Normalized parameters\tabularnewline
\hline 
\hline 
Time & $T_{\mathrm{p}}=\left(2\pi\right)/\left(R_{\mathrm{sc}}\omega_{\mathrm{p}}\right)$ & $\breve{\tau}_{\mathrm{w}}=\tau_{\mathrm{w}}/T_{\mathrm{p}}$\tabularnewline
\hline 
Length & $\lambda_{\mathrm{p}}=\left(2\pi c\right)/\left(R_{\mathrm{sc}}\omega_{\mathrm{p}}\right)$ & $\breve{z}=z/\lambda_{\mathrm{p}}$, $\breve{\sigma}_{\mathrm{B}}=\sigma_{\mathrm{B}}/\lambda_{\mathrm{p}}^{2}$\tabularnewline
\hline 
Frequency & $R_{\mathrm{sc}}\omega_{\mathrm{p}}=R_{\mathrm{sc}}\sqrt{4\pi\mathring{n}e^{2}/m}$ & $\check{\omega}=\omega/\left(R_{\mathrm{sc}}\omega_{\mathrm{p}}\right)$,
$\check{\omega}_{0}=\omega_{0}/\left(R_{\mathrm{sc}}\omega_{\mathrm{p}}\right)$\tabularnewline
\hline 
Velocity & $\mathring{v}$ & $\check{w}=w/\mathring{v}$, $\check{u}=u/\mathring{v}$, $\check{\gamma}=\gamma/\mathring{v}^{2}$,
$\breve{\beta}=\beta/\mathring{v}^{2}$\tabularnewline
\hline 
Wavenumber & $k_{\mathrm{p}}=2\pi/\lambda_{\mathrm{p}}=R_{\mathrm{sc}}\omega_{\mathrm{p}}/c$ & $\breve{k}=k/k_{\mathrm{p}}$\tabularnewline
\hline 
Charge & $Q_{0}\left(\omega=\omega_{0}\right)=A\tau_{\mathrm{w}}/\left(2\sqrt{\pi}\right)$ & $\breve{Q}=Q/Q_{0}\left(\omega=\omega_{0}\right)$\tabularnewline
\hline 
\end{tabular}
\end{table}


\end{document}